\documentclass[reprint,amsmath,amssymb,aps,prb,superscriptaddress,longbibliography]{revtex4-1}

\pdfoutput=1

\usepackage{hyperref,url}
\usepackage{amsmath}
\usepackage{amsfonts}
\usepackage{mathtools}
\usepackage[capitalise]{cleveref}
\usepackage{placeins}
\usepackage[version=4]{mhchem} 
\hypersetup{breaklinks,colorlinks=true,linkcolor=blue,citecolor=blue,urlcolor=blue,filecolor=blue}
\usepackage{graphicx}
\usepackage{dcolumn}
\usepackage{rotating}
\usepackage{etoolbox} 
\usepackage{multirow}
\newcommand{\beginsupplement}{%
        \setcounter{table}{0}
        \renewcommand{\thetable}{A\arabic{table}}%
        \setcounter{figure}{0}
        \renewcommand{\thefigure}{A\arabic{figure}}%
        \setcounter{equation}{0}
        \renewcommand{\theequation}{A\arabic{equation}}%
        \setcounter{section}{0}
        \renewcommand{\thesection}{A\arabic{section}}%
}

\newcommand\mat\mathbf

\newcommand{\Columbia}{\affiliation{Department of Chemistry, Columbia University, 3000 Broadway, New York, NY, 10027}}

\begin{document}

 \author{Joonho Lee}
 \email{linusjoonho@gmail.com}
 \Columbia
 \author{Hung Q. Pham}
 \Columbia
 \author{David R. Reichman}
 \Columbia
\title{Twenty Years of Auxiliary-Field Quantum Monte Carlo in Quantum Chemistry: An Overview and Assessment on Main Group Chemistry and Bond-Breaking}
\begin{abstract}
In this work,
we present
an overview of the phaseless auxiliary-field quantum Monte Carlo (ph-AFQMC) approach
from a computational quantum chemistry perspective, and present a numerical assessment of its performance on
main group chemistry and bond-breaking problems
with a total of 1004 relative energies.
While our benchmark study is somewhat limited, 
we make recommendations for the use of ph-AFQMC for general main-group chemistry applications.
For systems where single determinant wave functions are qualitatively accurate,
we expect the accuracy of ph-AFQMC in conjunction with a single determinant trial wave function to be between that of coupled-cluster with singles and doubles (CCSD) and CCSD with perturbative triples (CCSD(T)). For these applications, ph-AFQMC should be a method of choice when canonical CCSD(T) is too expensive to run.  For systems where multi-reference (MR) wave functions are needed for qualitative accuracy,
ph-AFQMC is far more accurate than MR perturbation theory methods and competitive
with MR configuration interaction (MRCI) methods. 
Due to the computational efficiency of ph-AFQMC compared to MRCI,
we recommended ph-AFQMC as a method of choice for handling dynamic correlation in MR problems.
We conclude with a discussion of important directions for future development of the ph-AFQMC approach. 
\end{abstract}
\maketitle
\newpage
\section{Introduction}
As computational power grows and theoretical algorithms have improved, the importance of the role played by electronic structure theory in chemistry, physics and biology has increased dramatically.\cite{helgaker2014molecular} For a number of important applications, the predictive power of quantum chemistry has reached the level of ``chemical accuracy,'' namely relative errors of one kcal/mol.~\cite{Mardirossian2017} However the most accurate methods are generally the most computationally expensive, and for systems with many electrons, often one must resort to approaches where convergence to predictive accuracy is far from guaranteed. When ``gold-standard'' wave function-based approaches are impractical, Kohn-Sham density functional theory (DFT) has been the method of choice, due to its combination of accuracy and mild scaling with system size.  However, there are many variants of DFT which depend on the choice of functional and it can be difficult {\em a priori} to select the most accurate density functional for a given application.  In addition, DFT generally suffers from self-interaction errors and has difficulty in the treatment of the effects of strong static correlation prevalent in the breaking of chemical bonds and in the electronic structure of transition metal-containing systems.~\cite{Cohen2012Jan} Indeed, even more expensive wave function approaches such as coupled-cluster theory\cite{bartlett_rmp} can have difficulty accurately describing strongly correlated cases.  Clearly there is still room for the development of novel methods in quantum chemistry which 
have a high ratio of accuracy to expense and may be reliably applied to challenging chemical problems.

Since the viability of an electronic structure method depends critically on the trade-off between accuracy and scaling with respect to system size, the development of new techniques that push the envelope of accuracy {\em and} scalability is the primary goal of the field.  Adoption of a method by the community cannot occur before careful and extensive benchmark studies are carried out to establish the domain of validity as well as the benefits and drawbacks of the method in question. It is our aim here to review the foundations of one particularly promising electronic structure framework, namely the auxiliary-field quantum Monte Carlo (AFQMC) approach, and to present, for the first time, the performance of this approach on a large-scale chemical benchmark set. 

Unlike DFT or conventional wave function methods, AFQMC employs a statistical as opposed to deterministic route to calculating electronic structure.  This necessitates consideration of an entirely different approach to the extraction of ground state energies, and requires the consideration of the role played by statistical fluctuations and error in the solution of the Schr{\"o}dinger equation.
The use and development of AFQMC has notably accelerated in recent years due to new algorithmic advances\cite{Malone2019Jan,Motta2019Jun,Lee2020Jul,Weber2022Jun,Mahajan2022May} and the publication of promising results on non-trivial chemical problems.\cite{al2006auxiliary,Purwanto2015Feb,Purwanto2016Jun,Shee2019Apr,Rudshteyn2020May,williams2020direct,Lee2020May,Rudshteyn2022May} Unlike standard DFT approaches, AFQMC is free of self-interaction error\cite{Lee2020May} (as it is a wavefunction method) and is often capable of describing strong correlation effects,\cite{Al-Saidi2006Jun,suewattana2007phaseless,Al-Saidi2007May,Al-Saidi2007Oct,Purwanto2009Mar,Purwanto2008Mar,Purwanto2011Oct,Motta2017Nov,Shee2017Jun,Motta2018May,Shee2018Aug,Shee2019Sep,LandinezBorda2019Feb,Lee2020Sep,Mahajan2022May} although our focus is not limited to such problems in this work.  

AFQMC has its roots in the determinant QMC (DQMC) method proposed by Blankenbecler, Scalapino, and Sugar.~\cite{Blankenbecler1981Oct}  DQMC is an exact, unbiased technique, initially developed to describe interacting bosons and fermions at finite temperatures in lattice field theory models of high energy physics. Since its invention, DQMC has been applied to numerous condensed matter problems, most notably the Hubbard model.~\cite{Hirsch1985Apr} More recently this approach has been applied to more realistic chemical systems with long-ranged interactions.~\cite{Silvestrelli1993Aug,Charutz1995Mar,Rom1998Nov,Mahajan2021Aug}  As is the case for DQMC, the scalability of AFQMC is limited by the fermionic sign problem, rendering exact, brute force applications to large chemical systems computationally infeasible.~\cite{Troyer2005May} To be more precise, the sign problem manifests in the following way: Per statistical sample, the complexity of the calculation scales polynomially with system size. However, the sample complexity (i.e., the number of statistical samples required for a fixed statistical error) scales  exponentially with system size, rendering
practical calculations impossible.  

To remedy this fundamental barrier in DQMC,  Fahy and Hamann proposed the approximate positive projection method,~\cite{Fahy1990Dec} which imposes a constraint to control the sign problem, resembling the fixed-node approximations used in Green's function Monte Carlo (GFMC)~\cite{Moskowitz1982Jul,vanBemmel1994Apr} and diffusion Monte Carlo (DMC).~\cite{Ceperley1980Aug}
Adopting this approximation for open-ended random walks, Zhang, Carlson, and Gubernatis~\cite{Zhang1995May}
proposed the constrained path approximation, and this method will be referred to as cp-AFQMC. 
cp-AFQMC has been successfully applied to non-trivial electronic lattice problems,~\cite{Zhang1997Jun,Guerrero1998May,Enjalran2001Oct,Chang2008Oct,Shi2013Sep,SimonsCollaborationontheMany-ElectronProblem2015Dec,Qin2016Aug,chang2016auxiliary,Zheng2017Dec,Qin2017Aug,Rosenberg2017Dec,Chiciak2018Jun,Vitali2019Apr,TheSimonsCollaborationontheMany-ElectronProblem2020Jul,SimonsCollaborationontheMany-ElectronProblem2020Jul,Chiciak2020Dec,Xu2022Mar,Vitali2022May} bosonic lattice problems,~\cite{Purwanto2004Nov,Purwanto2005Nov} and mixed fermions-bosons systems.~\cite{Rubenstein2012Nov,Lee2021Mar} 
cp-AFQMC has been shown to be one of the more accurate and scalable many-body methods in numerous benchmark studies,~\cite{SimonsCollaborationontheMany-ElectronProblem2015Dec,TheSimonsCollaborationontheMany-ElectronProblem2020Jul,SimonsCollaborationontheMany-ElectronProblem2020Jul} although it is important to bear in mind that like all constrained quantum Monte Carlo methods, cp-AFQMC is approximate and its domain of validity has not been fully mapped out.

The generalization of cp-AFQMC to chemical systems with long-range interactions
yields the phaseless AFQMC (ph-AFQMC) which is the focus of this work.~\cite{Zhang2003Apr}  Since its invention by Zhang and Krakauer in 2003,~\cite{Zhang2003Apr} ph-AFQMC has gained popularity in the quantum chemistry community due to its relative computational affordability and its accuracy, as demonstrated in recent calculations including those performed on small but challenging molecular systems,~\cite{Al-Saidi2006Jun,suewattana2007phaseless,Al-Saidi2007May,Al-Saidi2007Oct,Purwanto2009Mar,Purwanto2008Mar,Purwanto2011Oct,Motta2017Nov,Shee2017Jun,Motta2018May,Shee2018Aug,Shee2019Sep,LandinezBorda2019Feb,Lee2020Sep,Mahajan2022May}
simple transition metal complexes,~\cite{al2006auxiliary,Purwanto2015Feb,Purwanto2016Jun,Shee2019Apr,Rudshteyn2020May,williams2020direct,Lee2020May,Rudshteyn2022May}
and solids.~\cite{purwanto2009pressure,ma2015quantum,motta2017towards,ma2017auxiliary,Zhang2018Oct,Eskridge2019Jul,Malone2019Jan,Lee2019Aug,motta2019hamiltonian,motta2020ground,Malone2020Jul,Morales2020Nov,malone2020systematic,chen2021ab,Lee2021Jun}  ph-AFQMC is highly flexible in that the accuracy of the approach may be systematically improved by the use of increasingly sophisticated trial wave functions.  In addition, ph-AFQMC has become more accessible to users via the development of open-source codes.~\cite{Kent2020May,ipie,Malone2022Sep}

Despite its growing  usage in the chemistry community,
it is not clear how well ph-AFQMC performs for a wide range of main group chemistry applications.
While it has been speculated that ph-AFQMC with the simplest form of a single-reference trial wave function has an accuracy on par with coupled-cluster with singles, doubles, and perturbative triples (CCSD(T)),~\cite{Raghavachari1989May,Al-Saidi2006Jun,Lee2019Aug,Lee2020Sep} 
there has been no extensive body of work that confirms this expectation.
While ph-AFQMC has been compared to other commonly used quantum chemistry methods such as CCSD(T) on simple bond breaking problems 
such as those that occur in \ce{H2O},~\cite{Al-Saidi2006Jun} \ce{BH},~\cite{Al-Saidi2007Oct} \ce{N2},~\cite{Al-Saidi2007Oct} \ce{F2},~\cite{Purwanto2008Mar} \ce{C2},~\cite{Purwanto2009Mar} \ce{Cr2},~\cite{Purwanto2015Feb} and ~\ce{Mo2},~\cite{Purwanto2016Jun} studies comparing ph-AFQMC with both single-reference CCSD(T) and other multi-reference methods have been relatively scarce.
Given the accuracy of ph-AFQMC for dynamic correlation problems,~\cite{motta2017towards,motta2020ground,williams2020direct,Lee2020Sep}
it is crucial to compare its accuracy with other multi-reference dynamic correlation methods, such as perturbation theory and configuration interaction methods.

The goal of this manuscript is both to present a pedagogical review of ph-AFQMC from the quantum chemistry perspective as well as to investigate the performance of ph-AFQMC on a large chemical benchmark series, namely a thermochemistry benchmark set (W4-11),~\cite{KARTON2011165} a non-covalent interaction benchmark set (A24),~\cite{rezac2013describing} and model molecular bond dissociation problems (\ce{H4} and \ce{N2}).~\cite{Jankowski1980Nov,Siegbahn1983Jun,paldus1993application,Mahapatra1999Apr,kowalski2000complete,Evangelista2006Oct,Ma2006Jan,Small2012Sep,Lee2019Jan}  From this, we hope to establish fair expectations for ph-AFQMC for main group chemistry and simple strong correlation problems associated with the breaking of chemical bonds,
and to further encourage the use and development of ph-AFQMC in quantum chemistry in the future. 
We note that there exist other pedagogical reviews written from a somewhat different perspective~\cite{Zhang2013,Motta2018Sep,Shi2021Jan} which the interested reader is encouraged to consult.

It should be noted that the applications considered here, while chemically important, are somewhat uncommon for ph-AFQMC.  In particular, the benchmark sets (W4-11~\cite{KARTON2011165} and A24~\cite{rezac2013describing}) considered here are predominantly single reference problems. These do not involve strong electron correlation for which ph-AFQMC has been frequently used. Indeed, these benchmark problems are already known to be well-characterized by coupled-cluster techniques. Furthermore, we mostly consider the {\em least} accurate and the most efficient variant of ph-AFQMC. Namely we employ a simple single-determinant trial wave function, although we do discuss selected examples where the use of multi-determinant trials can greatly improve accuracy.  These choices are purposeful ones: we aim to assess the breadth and baseline accuracy of the approach in situations where other quantum chemistry methods such as CCSD(T) are often well-suited. The main points that we will deliver in this manuscript are as follows:
\begin{enumerate}
\item For main group chemistry applications, with $\mathcal O(N^{3})$ -- $\mathcal O(N^4)$ scaling per sample, ph-AFQMC with a single-reference trial wave function
is more accurate than CCSD ($\mathcal O(N^6)$) and is competitive with, albeit somewhat less accurate, than CCSD(T) ($\mathcal O(N^7)$) and state-of-the-art density functionals.
\item For bond breaking, while maintaining $O(N^{3})$ -- $\mathcal O(N^4)$ scaling per sample, the accuracy of ph-AFQMC with a multi-reference trial wave function
surpasses that of low-order multi-reference perturbation theory ($\mathcal O(N^5)$ -- $\mathcal O(N^6)$) and is comparable in accuracy with multi-reference configuration interaction methods ($\mathcal O(N^6)$).
\item Given the relatively short history of ph-AFQMC in quantum chemistry,
many aspects of the algorithm deserve further
investigation and there is much room for improvements to the current implementation of the approach. We will conclude this work with a discussion of these opportunities.
\end{enumerate}

This manuscript is organized as follows: In \cref{sec:overview} we present an overview of the AFQMC method and its phaseless variant (ph-AFQMC). A discussion of trial wave functions, the calculation of observables, size-consistency and computational cost is contained in this section.  In \cref{sec:benchmark} we present benchmark results with a comparison with other approaches. \cref{sec:lessons} presents a discussion of lessons learned from our calculations and prospects for future development. We conclude in \cref{sec:conclusions}. The Appendices contain details omitted from the main text.

\section{Overview of phaseless auxiliary-field Quantum Monte Carlo}\label{sec:overview}

\subsection{Ground-state calculations}
We refer the interested readers to \cref{appsec:ground} for a more detailed formal exposition of AFQMC.
Here, we focus on a short summary of the methodology.
We assume the spin-orbital notation throughout this paper.

AFQMC is what is called a ``projector'' QMC algorithm because it projects towards the ground state from an initial wave function which has a non-zero overlap with the true, exact ground state of the system.  Formally, the ground-state wave function $|\Psi_{0}\rangle$ of a Hamiltonian $\hat{H}$ is found by via imaginary time propagation
\begin{equation}
|\Psi_{0}\rangle = \lim_{\tau \to \infty} e^{-\tau\hat{H}}|\Phi_{0}\rangle,
\label{eq:imag_proj}
\end{equation}
where $\hat{H}$ is an {\it ab initio} Hamiltonian,
\begin{equation}
\hat{H}
=
\sum_{pq}h_{pq}a_p^\dagger a_q
+ \frac12 \sum_{pqrs}
(pr|qs) a_p^\dagger a_q^\dagger a_s a_r,
\label{eq:Habinitio}
\end{equation}
$\tau$ is the imaginary time, and
$|\Phi_0\rangle$ is an initial wave function that is not orthogonal to $|\Psi_{0}\rangle$. 
The form of $|\Psi_{0}\rangle$ in \cref{eq:imag_proj} resembles
the CCSD wave function with generalized singles and doubles,\cite{nakatsuji1976equation,nooijen2000can} but
in AFQMC there is no cluster amplitude to be determined, and formally~\cref{eq:imag_proj} is exact.

Discretizing $\tau$ and applying the Trotter decomposition~\cite{Trotter1959Aug} followed by the Hubbard-Stratonovich (HS)~\cite{hubbard_strat} transformation to the propagator $e^{-\Delta \tau\hat{H}}$, one obtains a high-dimensional integral formula for the imaginary time evolution operator over a short time interval
\begin{equation}
e^{-\Delta \tau\hat{H}} = \int \mathrm{d}^{N_\gamma}\textbf{x} ~p(\textbf{x}) \hat{B}(\textbf{x}) + \mathcal O (\Delta \tau^2),
\label{eq:Bx}
\end{equation}
where $\Delta\tau$ is the infinitesimal time step, $p(\textbf{x})$ is the standard normal distribution and $\hat{B}(\textbf{x})$ is
an effective one-body propagator that is obtained from one-body operators coupled to 
\textbf{x}, a vector of $N_{\gamma}$ classical auxiliary fields. Since only one-body operators now appear, this representation of the short-time propagator is often viewed as the exact mapping of an interacting many-body system to an ensemble of non-interacting systems, each coupled to a classical, fluctuating external potential. 
The many-body correlation beyond mean-field theory is recovered upon the integration over the auxiliary fields.

The phaseless AFQMC (ph-AFQMC)\cite{Zhang2003Apr}  approach utilizes importance sampling based on a trial wave function $|\Psi_T\rangle$   during the random walk process.  To carry this out, one writes the global wave function at time $\tau$ as a weighted statistical sum over $N_\text{walkers}$ ``walkers'',
\begin{equation}
|\Psi(\tau)\rangle = 
\sum_{i=1}^{N_\text{walkers}}
w_i (\tau)
\frac{|\psi_i(\tau)\rangle}
{\langle \Psi_T|\psi_i(\tau)\rangle}.
\label{eq:stat_wfs}
\end{equation}
Given this importance sampling,
the walker weight update rule follows
\begin{equation}
w_{i}(\tau+\Delta\tau) = 
w_i(\tau) \times
S ( \textbf{x}_i(\tau),  |\psi_{i}{(\tau)}\rangle ), 
\label{eq:import_sampling}
\end{equation}
where the overlap ratio $S$ is given by
\begin{equation}
S ( \textbf{x}_i(\tau), |\psi_{i}{(\tau)}\rangle ) =
\frac{\langle \Psi_{{T}} | \hat{B}(\textbf{x}_i(\tau)) |\psi_{i}{(\tau)} \rangle}{ \langle \Psi_{{T}} |\psi_{i}(\tau) \rangle }.
\label{eq:overlap}
\end{equation}
Since this weight update rule cannot guarantee that the walker weights are real and positive, $\{w_i\}$,
\cref{eq:import_sampling} leads to the fermionic phase (sign) problem
with a diverging variance for any observable when calculated by the protocol outlined below.~\cite{Troyer2005May}

To control the phase problem, ph-AFQMC imposes a constraint to ensure the positivity of the weights throughout the imaginary time propagation.  The constraint that defines the standard ph-AFQMC approach is given by
a modified overlap ratio, $S_{\text{ph}}$,
\begin{align}\nonumber
S_{\text{ph}} ( \textbf{x}_i(\tau),|\psi_{i}{(\tau)}\rangle ) 
&= 
||S ( \textbf{x}_i(\tau)|\psi_{i}{(\tau)}\rangle )||\\
&\times \text{max}\left(0, \cos \theta_i (\tau) \right),
\label{eq:mod_overlap}
\end{align}
where the phase $\theta_i(\tau)$ is given by
\begin{equation}
\theta_i (\tau) = \text{arg} ~S ( \textbf{x}_i(\tau)|\psi_{i}{(\tau)}\rangle ).
\label{eq:phase}
\end{equation}
This modified overlap ratio is used in ph-AFQMC to update the weights.
It should be emphasized that this constrained random walk introduces systematic biases which induce deviations in the values of observables (such at the ground state energy) compared to the exact values. It should be noted that this bias disappears in the limit of
$|\Psi_T\rangle$ becoming an exact eigenstate of $\hat{H}$.  This fact introduces a means to improve the accuracy of ph-AFQMC by increasing the sophistication of the trial function $|\Psi_T\rangle$.  We will return to this point later in this work.
A different way to view the constraint is that one imposes a gauge boundary condition (i.e., a global phase)
for the wave function sampled through the imaginary time evolution. 

Within this framework,
the global energy estimate at a given imaginary time $\tau$ is given by a weighted statistical average,
\begin{equation}
\langle O(\tau)\rangle_\text{mixed} = 
\frac
{
\sum_i w_i(\tau) O_{L,i}(\tau)
}
{\sum_i w_i(\tau)},
\label{eq:Em2}
\end{equation}
where the local estimate  for observable $\hat{O}$ of the $i$-th walker is given by
\begin{equation}
O_{L,i}(\tau)
=
\frac{\langle \Psi_T | \hat{O} | \psi_i(\tau)\rangle}
{\langle \Psi_T |  \psi_i(\tau)\rangle}.
\label{eq:elocal}
\end{equation}
If $\hat{O} = \hat{H}$, these estimates are global energy and local energy estimates, respectively.
Given the positivity of $\{w_i\}$, the variance of this energy estimate 
grows linearly with system size, ensuring the polynomial-scaling sample complexity of the overall algorithm.
It is important to remark here that the resulting ph-AFQMC energy computed via \cref{eq:Em2} is not guaranteed to be variational.~\cite{carlson_no_var}

\subsection{Trial wave functions} \label{subsec:trial}
The accuracy of  ph-AFQMC heavily depends
on the quality of the trial wave function employed in the calculation.
Conceptually, the role played by the trial wave function is very different from that of the reference wave function
used in conventional quantum chemistry methods such as CCSD(T).  Despite this difference, it is convenient to think of ph-AFQMC as adding correlation energy on top of an {\it a priori} chosen trial wave function.
Here, we summarize existing strategies for generating these ph-AFQMC trial states.

Single determinant (SD) trial wave functions are the most widely used and they offer the most affordable variant of the ph-AFQMC algorithm.  There are multiple approaches to obtain single determinant trial wave functions.
One way is to employ the lowest energy spin-unrestricted or spin-generalized Hartree-Fock (HF) wave function.~\cite{LandinezBorda2019Feb,Lee2020May} This approach can be well-suited for describing bond dissociation as shown in \cref{sub:bondbreaking}.
While this approach is completely parameter-free, it often runs into issues associated with artificial symmetry breaking.~\cite{Lee2019Feb,Lee2020May}  In particular, HF wave functions can exhibit an unphysical breaking of symmetry that makes post-HF calculations behave erratically in energetics and properties.~\cite{Farnell1983,Nobes1987,Gill1988,Jensen1990,Andrews1991,Yamanaka1994,Ayala1998,Crawford2000}
While ph-AFQMC can correct artificial symmetry breaking to some extent through imaginary time evolution, 
there are cases where the lowest energy HF solution is clearly not the best SD trial wave function.\cite{Lee2020May}  Alternatively, one can obtain a single Slater determinant from methods that include electron correlation effects.  For this, DFT~\cite{Zhang2003Apr,al2006auxiliary} and regularized orbital optimized M{\o}ller-Plesset perturbation theory (OOMP2) have been used.~\cite{Lee2018Oct,Lee2020May}  More broadly, one can employ any approximate Brueckner orbital wave functions ~\cite{Dykstra1977Feb,Raghavachari1990Mar} for this purpose. 
We also note that the recently-defined self-consistent trial wave function method~\cite{qin2016coupling,Shi2021Jan} also fits into the category of an SD trial wave function approach.

More elaborate trial wave functions may be used within ph-AFQMC, which confer greater accuracy for a greater computational expense. Most commonly, a linear combination of SDs, often referred to as multi-Slater determinants or MSDs for short, are employed.~\cite{Al-Saidi2007Oct,Purwanto2008Mar,LandinezBorda2019Feb,Mahajan2022May} 
The MSD trial can be obtained from a truncated configuration interaction (CI) expansion, complete active space CI (CASCI), 
or from selected CI methods.~\cite{Holmes2016Aug}
While these trial wave functions can yield excellent ph-AFQMC energies, they are brute-force in nature and the cost of obtaining these trial wave functions generally scales exponentially with system size, which can limit the applicability of their use within ph-AFQMC.

Lastly, there are non-linear trial wave functions which may be used. These include Jastrow factors,~\cite{hlubina1997ferromagnetism,chang2016auxiliary} coupled-cluster wave functions,~\cite{bartlett_rmp,Huggins2022Mar,Mahajan2021Aug} perfect-pairing wave functions,~\cite{Goddard1973Nov,Purwanto2008Mar,Huggins2022Mar} and transcorrelated wave functions.~\cite{Schraivogel2021Nov}
The cost of obtaining these wave functions in all such cases is polynomial-scaling, which makes them appealing for use as trials. Furthermore, these wave functions approximate the exact ground state much more accurately than do SD trial functions
as they all include electron correlation inherently missing from the SD description.
Unfortunately, using these non-linear wave function as a trial wave functions without any approximations is
currently extremely difficult on both classical and quantum computing platforms.~\cite{Huggins2022Mar} The efficient and accurate use of such trial states within ph-AFQMC will require further work to become viable.

\subsection{Physical properties extracted from ph-AFQMC}
In practical applications, one may want to evaluate observables that do not
commute with $\hat{H}$. Most commonly, these observables are associated directly with reduced density matrices (RDMs) or some elements of them.
The computation of such observables introduces additional challenges within the  ph-AFQMC framework,
because the mixed estimator in \cref{eq:Em2} is no longer unbiased.
For such quantities, we need to use a different estimator known as the ``pure'' estimator.
This problem is analogous to the issues associated with coupled-cluster expectation values~\cite{Bartlett1988Sep}
which are typically now handled by the coupled-cluster Lagrangian formalism.~\cite{Pedersen1997May} 

A computationally simple approach to this problem is to
use the approximate variational estimator
\begin{equation}
\langle O(\tau) \rangle_\text{pure}
\approx
2 \langle O(\tau) \rangle_\text{mixed} - \frac{\langle \Psi_T | \hat{O} | \Psi_T \rangle}{\langle \Psi_T | \Psi_T \rangle},
\label{eq:pure1}
\end{equation}
which is frequently used in diffusion Monte Carlo.~\cite{whitlock1979properties,RothsteinStuart2013May}
Recently, \cref{eq:pure1} was used within the ph-AFQMC framework to estimate the dipole moments of simple molecules.~\cite{Mahajan2022May} Here, it was found that the quality of the trial wave function is critically important for accurately approximating the pure estimates.  For this quantity, using SD trial wave functions yielded very inaccurate results for the systems considered in Ref. ~\citenum{Mahajan2022May}.

Another approach that is more commonly used in ph-AFQMC calculations is the back-propagation algorithm.~\cite{zhang1997constrained,Purwanto2004Nov,Motta2017Nov}
Here, one propagates the bra state ($\langle \Psi_T|$) in the mixed estimator (\cref{eq:Em}) backward in time
such that
the resulting estimator is symmetric with respect to both the
bra and ket, ultimately resembling the pure (or variational) estimator.
Formally, this entails computing observables using the following estimate:
\begin{align}\nonumber
\langle {O(\tau)} &\rangle_\text{pure} 
\approx  
\lim_{\kappa\rightarrow\infty} 
\frac{\langle \Psi_T | e^{-\kappa \hat{\mathcal{H}}} \hat{O} | \Psi(\tau)\rangle}
{\langle \Psi_T | e^{-\kappa \hat{\mathcal{H}}} | \Psi(\tau)\rangle}
\\
&=
\lim_{\kappa\rightarrow\infty} 
\frac{
\sum_i w_i(\tau+\kappa)
    \frac{
        \langle \psi_i(\kappa) | \hat{O} | \psi_i(\tau) \rangle
    }
    {
        \langle \psi_i(\kappa) | \psi_i(\tau)\rangle
    }
}
{
    \sum_i w_i(\tau+\kappa) ,
}\label{eq:bp_obs}
\end{align}
where the back-propagation time, $\kappa$, while formally taken to $\infty$, is in practice fixed to a long enough finite time length.  This approach is computationally efficient, but its accuracy for at least some {\it ab initio} systems has been shown to be 
rather poor.~\cite{Motta2017Nov, Lee2021Jun}
There are additional algorithmic ways to reduce the back-propagation bias, such as partially restoring the phase and cosine factors along the back propagation portion of the path.~\cite{Motta2017Nov, Shi2021Jan}
However, even with these considerations, the resulting 1-RDMs were found to be inaccurate, at least in some systems.~\cite{Lee2021Jun}  Devising improved algorithms for accurately computing pure estimates for observables without greatly increasing the computational cost of the calculations is a worthy goal for future research.

\subsection{Size-consistency}\label{subsec:size}
Size-consistency is a property of a wave function for isolated systems $A$ and $B$ that guarantees the product separability of a supersystem
wave function ($|\Psi_{AB}\rangle = |\Psi_A\rangle|\Psi_B\rangle$) as well as the additive separability of energy ($E_{AB} = E_A+E_B$).~\cite{Bartlett1981Oct}
This property has important implications concerning the applicability of a given method to large systems
and is therefore considered to be an important formal property in method development.
The size-consistency of ph-AFQMC was first examined in Ref. \citenum{Lee2019Aug}, but some subtle issues were not fully discussed in that work. Here, we will provide the first rigorous analysis of size-consistency within ph-AFQMC.

We will assume that the trial wave function is product separable, and that for isolated systems $A$ and $B$, the total Hamiltonian separates into $\hat{H}_A$ and $\hat{H}_B$ which commute with each other.  Given these conditions, the propagator is product separable,
\begin{equation}
\exp(-\Delta\tau \hat{H}_{AB})
=
\exp(-\Delta\tau \hat{H}_{A})
\exp(-\Delta\tau \hat{H}_{B}),
\end{equation}
which leads to the 
product separability of $\hat{B}$, $\hat{B}_{AB} = \hat{B}_A\hat{B}_B$.

Provided that the walker wave function is product separable, it can be shown that the overlap ratio in~\cref{eq:overlap} of the total system, $S^{AB}$, is product separable into monomer overlap ratios, $S^A$ and $S^B$
\begin{equation}
S^{AB} =
\frac{\langle \Psi_{{T}}^A | \hat{B}_A |\psi^A \rangle}{ \langle \Psi_{{T}} |\psi^A\rangle }\frac{\langle \Psi_{{T}}^B | \hat{B}_B |\psi^B \rangle}{ \langle \Psi_{{T}} |\psi^B\rangle }
=
S^A S^B.
\end{equation}
The product separability of the walker wave function can be satisfied as long as we start from a product separable wave function since the propagator itself is product separable.
With this overlap ratio, walker weights are also product separable as $w^{AB} = w^A w^B$.
Since the local energy in~\cref{eq:elocal} is additively separable ($E_L^{AB} = E_L^A + E_L^B$),
we conclude that $\langle E^{AB}\rangle = \langle E^A \rangle + \langle E^B \rangle$.

However, the above analysis does not apply to the modified overlap ratio for arbitrary imaginary time steps in \cref{eq:mod_overlap}
used in the constrained ph-AFQMC formalism.  To see this, noting the cosine projection which defines the constraint for the ph-AFQMC framework, we have
\begin{align}
\exp(i\theta^{AB}) &= \exp(i\theta^{A}) \exp(i\theta^{B})\\
\label{eq:cosadd}
\cos(\theta^A + \theta^B)
&=
\cos(\theta^A)\cos(\theta^B)
-\sin(\theta^A)\sin(\theta^B),
\end{align}
where
the cosine factor is not product separable, leading to size-inconsistency in the overall approach if the time step is not taken to zero.
In other words, the 
walker weights of the total system are not
product separable ($w^{AB} \neq w^A w^B$).
The magnitude of the size-inconsistency error is proportional to the magnitude of the sine terms in \cref{eq:cosadd}, and can be practically small.  Importantly, if the time step $\Delta t$ is small, then $\theta^A$ and $\theta^B$ are small. Therefore,
we expect the size-inconsistency error to vanish in the limit of $\Delta t \rightarrow 0$.  We present numerical results to support this in~\cref{appsubsec:size}.
In this sense, the size-inconsistency error is a part of the time step error with an $\mathcal O(\Delta t)$ error scaling as opposed to $\mathcal O(\Delta t^2)$ of the Trotter error.
Like other time step errors, the above analysis suggests that size-inconsistency error can be controlled within ph-AFQMC by taking the limit of $\Delta t \rightarrow 0$. 
While we examined the size-consistency of ph-AFQMC assuming localized orbitals and product-separability of the trial wavefunction, it is possible that these assumptions may not be necessary to examine the size-consistency more generally.

In summary, we have shown that ph-AFQMC is strictly size-consistent in the limit of $\Delta t\rightarrow 0$. The size-consistency is critically important to reliably apply ph-AFQMC to large systems and solid state problems, similarly to more traditional quantum chemistry methods such as coupled-cluster theory.

\subsection{Computational cost}
\begin{table}[h!]
\begin{tabular}{|c|c|c|c|}
\hline
             & SD & MSD & Non-linear \\ \hline
$\langle \Psi_T | \psi_i (\tau)\rangle$  &  $\mathcal O(N^3)$ &  $\mathcal O(N_c+N^3)$\cite{Mahajan2021Aug}   &    $\mathcal O(e^N)$        \\ \hline
Green's function & $\mathcal O(N^3)$ &   $\mathcal O(N_c+N^3)$\cite{Mahajan2022May}  &     $\mathcal O(e^N)$       \\ \hline
\multirow{2}{*}{Local energy} & $\mathcal O(N^4)$ &   $\mathcal O(N_c N+N^4)$\cite{Mahajan2021Aug}   & \multirow{2}{*}{$\mathcal O(e^N)$}           \\ 
& $\mathcal O(N^3)$ &$\mathcal O(N_cN^2+N^3)$\cite{Weber2022Jun} &\\\hline
\end{tabular}
\caption{Summary of ph-AFQMC per-sample costs of each component (overlap, one-body Green's function (see \cref{eq:1gf}), and local energy (see \cref{eq:elocal})) for single-determinant (SD), multi Slater determinant (MSD), and non-linear trial wave functions. 
$N$ denotes system size and $N_c$ is the number of determinants in an MSD trial wave function.
The cubic-scaling for the local energy evaluation in the case of SD follows from arguments presented in
Refs.
~\citenum{Motta2019Jun,Malone2019Jan,Lee2020Jul,Weber2022Jun}.
}
\label{tab:cost}
\end{table}

The computational expense and scaling with system size of ph-AFQMC is an important factor when considering suitable applications for this approach.
The integral transform often necessary for ph-AFQMC scales as $\mathcal O(N^4)$.
There are three additional considerations that need to be considered within ph-AFQMC to determine the scaling  behavior and how it depends on the choice of a trial wave function.
For the following discussion, we will assume that the walker wave functions are SDs.
The first consideration concerns the propagation step.
The central quantity to be computed is the modified overlap ratio in~\cref{eq:mod_overlap}.  Specifically, the overlap between the trial and walker wave functions, $\langle \Psi_T | \psi_i (\tau)\rangle$, must be
efficiently evaluated if the method is to be computationally viable.  
In addition to the overlap, one needs to evaluate the one-body Green's function for the force bias evaluation (see ~\cref{appsec:ground} for more details).
The final consideration relates to cost of the local energy evaluation in~\cref{eq:elocal} assuming that the ground state energy 
is the quantity of interest.
In~\cref{tab:cost} we summarize the cost of these three parts for different types of trial wave functions discussed in~\cref{subsec:trial}.

For the local energy evaluation there has been a variety of recent algorithmic improvements.
For example, the best algorithms now available reduce the standard quartic-scaling algorithm of the SD local energy evaluation to a cubic-scaling algorithm.  This can be achieved by using
double factorization,~\cite{Motta2019Jun} tensor hypercontraction,~\cite{Malone2019Jan}
stochastic resolution-of-the-identity,~\cite{Lee2020Jul}
or via the use of localized orbitals.~\cite{Weber2022Jun}
Accelerating the local energy evaluation with multi-SD trials has not been explored as extensively as that of the simpler SD case, but recent explorations with localized orbitals~\cite{Weber2022Jun} and with generalized Wick's theorem\cite{Mahajan2021Aug,Mahajan2022May} are encouraging.
Future work should be aimed at lowering the complexity of the overlap and the force bias evaluation 
since these computations often form the bottleneck for medium-sized molecules.

If one is to compare the cost of standard deterministic quantum chemistry methods to the cost of ph-AFQMC,
a subtlety arises due to the statistical nature of ph-AFQMC.
For bulk systems when energy per particle or other size-intensive quantities are relevant, our cost analysis given in Table \cref{tab:cost} is sufficient, assuming the number of samples required for desired precision does not grow with system size.
For finite molecular systems, it is sensible to estimate the cost of ph-AFQMC
for a fixed statistical error as the system size, $N$, increases.
Assuming that the auto-correlation time in the Markov chain does not grow with $N$
and that the standard error in the energy estimate grows linearly with $N$,
one crudely requires $\mathcal O(N^2)$ statistical samples to maintain a fixed statistical error.
Therefore, we find a
$\mathcal O(N^2)$ multiplicative factor
in addition to
the cost in~\cref{tab:cost}.
While this is a correct formal asymptotic scaling,
it is possible that in practice
one may not experience $\mathcal O(N^2)$  sample complexity
if the trial wave function is accurate such that
the statistical fluctuations are suppressed below a desired error threshold for the range of system sizes under consideration.

\section{Performance of ph-AFQMC for thermochemistry and the treatment of non-covalent interactions and bond dissociation}\label{sec:benchmark}
We will now assess the accuracy of ph-AFQMC using well-known thermochemistry and non-covalent interaction benchmark sets, along with several simple bond dissociation examples. For simplicity, we will refer to ph-AFQMC as AFQMC in this section. Computational details are available in~\cref{appsub:compdetails}.  Note that nearly all of the AFQMC calculations presented here are new and have not been previously published.

\subsection{Thermochemistry benchmark (W4-11)}
W4-11~\cite{KARTON2011165} is a high-quality benchmark set with a total of 979 relative energies. This data set has been extensively used to assess the performance of distinct density functionals~\cite{Goerigk2017Dec, Mardirossian2017} and other quantum chemical methods.~\cite{Jankowski1980Nov,Siegbahn1983Jun,paldus1993application,Mahapatra1999Apr,kowalski2000complete,Evangelista2006Oct,Ma2006Jan,Small2012Sep,Lee2019Jan} It covers a variety of chemical reactions for the main-group elements, including 140 total atomization energies (TAE140),  707 heavy-atom transfer energies (HAT707), 99 bond dissociation energies (BDE99), 20 isomerization energies (ISO20), and 13 nucleophilic substitution reaction energies (SN13). In addition, it is an economical benchmark set because
one can generate the remainder of the 839 data points using the TAE140 data, which only requires 152 single point energy calculations.

We note that not every data point in the W4-11 set is a simple single-reference (SR) problem.
The TAE140 subset contains 16 multi-reference (MR) data points determined by the \%TAE$_e$[(T)] diagnostic.~\cite{KARTON2011165} Similarly, 202 energies in the HAT707 set and 16 energies in the BDE99 set have also been deemed MR data points.~\cite{Mardirossian2017}
We will discuss the performance of ph-AFQMC on 745 SR data points seperately from the 234 MR data points, as is often done for density functionals and other quantum chemistry methods.~\cite{Mardirossian2017}

\subsubsection{Single atom total energies}\label{subsub:atom}

\begin{figure}[!h]
    \centering
    \scalebox{0.5}{\includegraphics{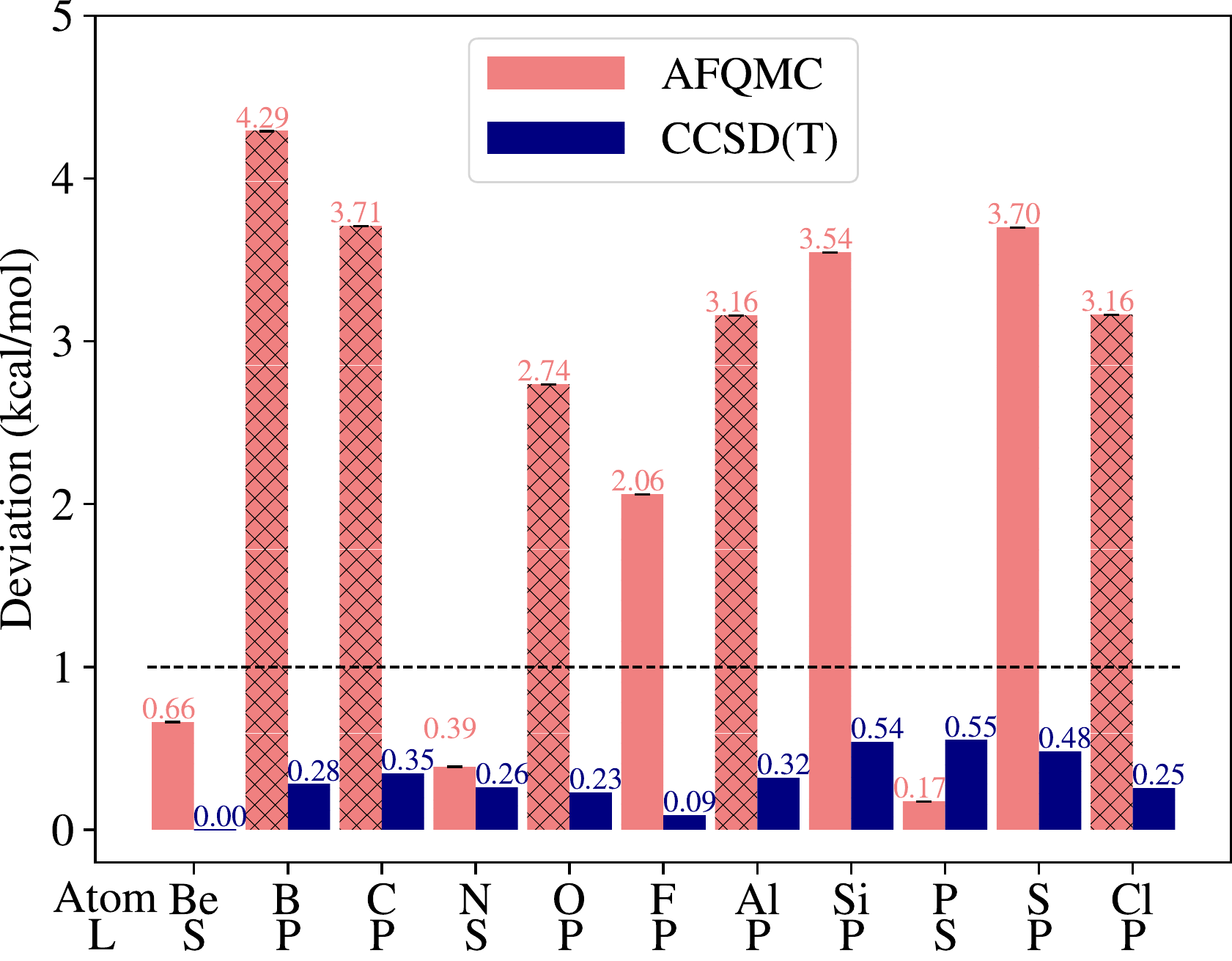}}
    \caption{Deviation (in kcal/mol) of AFQMC (blue) and CCSD(T) (orange) from exact energies in an aug-cc-pVTZ basis. The error bars for AFQMC are nearly undetectable on this energy scale.
    In the X-axis ticks, we also indicate the angular momentum (L) of the ground state of each atom.
    The black dotted lines indicate the ``chemical accuracy'' value of 1 kcal/mol.}
    \label{w411atoms}
\end{figure}

The simplest systems in the W4-11 set are single atoms (Be, B, C, N, O, F, Al, Si, P, S, Cl).
These are systems are small enough that brute-force methods such as SHCI can be reliably performed to obtain exact total energies. We first look at the performance of CCSD(T) and AFQMC for these in the aug-cc-pVTZ basis set as shown in \cref{w411atoms}.
We used spin-restricted open-shell Hartree-Fock (ROHF) as a reference state for CCSD(T) and as a trial state for AFQMC.  The conclusions are unchanged if one uses a spin-unrestricted Hartree-Fock (UHF) trial instead.

In~\cref{w411atoms}, CCSD(T) achieves chemical accuracy (1 kcal/mol) for all the atoms in W4-11.
However, AFQMC exhibits large errors above 2 kcal/mol for numerous atoms.
There is an interesting trend that the AFQMC error becomes an order of magnitude smaller for atoms whose ground state angular momentum is of S symmetry. For all the other cases (ground state angular momentum, P), AFQMC errors are greater than 2 kcal/mol.  We conducted a preliminary investigation of this trend using various different SD and MSD trials
with the hypothesis that these large errors can be attributed to simple symmetry constraints, as seen in the Hubbard model.~\cite{Shi2013Sep,Qin2016Aug} 
Unfortunately, while one can converge all of the atomic energies to chemical accuracy with a sophisticated multi-SD trial, we were unable to find a simple, compact trial wave function that removes this bias. For instance, we investigated C in aug-cc-pVTZ further with an exact trial wavefunction generated by SHCI. To achieve chemical accuracy, we needed more than 200 determinants in the trial wavefunction which has more than 0.99 overlap with the exact ground state. 

Very recent work suggests that using SD trial wave functions with a partially relaxed constraint may reduce the bias significantly although the accuracy of such a constraint on large systems remains unclear.~\cite{weberunpublished}
Alternatively, one could even perform free-projection\cite{Mahajan2021Aug} or release constraint\cite{Shi2013Sep} calculations with SD trial wave functions on these small atoms, but this is not a viable option for larger systems.
We leave further investigation of these simple systems for future studies, and focus on the performance of AFQMC for W4-11 where we use CCSD(T) atomic energies but AFQMC molecular energies. Lastly, we note that due to the frozen core approximation, Be contains only two electrons that need to be correlated.  Therefore, CCSD is exact for this system, as shown in~\cref{w411atoms}. However, this exactness does not apply to AFQMC.

\subsubsection{Single-reference systems}
\begin{figure}[!h]
    \centering
    \scalebox{0.49}{\includegraphics{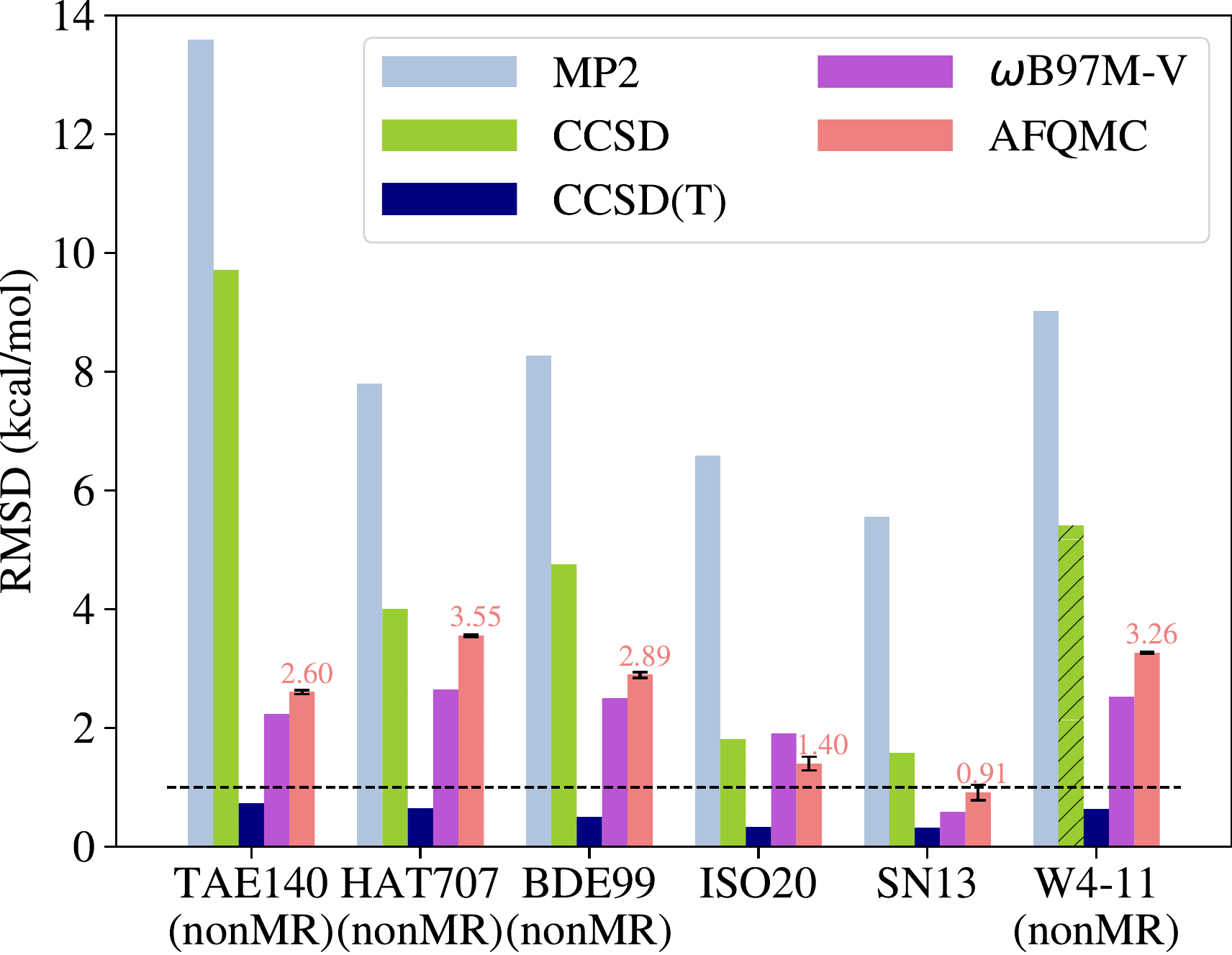}}
    \caption{Root-mean-square deviation (RMSD) for each subset of the W4-11 data set and for the overall W4-11 set, including only nonMR (i.e., SR) data points. The black dotted lines indicate the ``chemical accuracy'' value of 1 kcal/mol.}
    \label{fig:w411nonmr}
\end{figure}
We have performed CCSD(T) and AFQMC using both SD RHF and UHF states as references or trial functions, respectively.
We report the raw energies of each case in the Zenodo repository.\cite{zenodo}
For AFQMC, it was found that the use of UHF trials leads to more accurate results, while for CCSD(T) using an RHF references was statistically superior. Hence, we will compare AFQMC with UHF trials and CCSD(T) with RHF reference for all single-reference calculations.

In~\cref{fig:w411nonmr}, we observe systematic improvements for both the root-mean-square deviation (RMSD) and mean-signed error (MSE) as we increase the sophistication of the correlation treatment, with second-order M{\o}ller-Plesset perturbation theory (MP2) less accurate than CCSD which is less accurate than CCSD(T).
However, both MP2 (9.01 kcal/mol) and CCSD (5.40 kcal/mol) do not achieve chemical accuracy in terms of RMSD for any of the W4-11 subsets.
CCSD(T) achieves chemical accuracy in all subsets of W4-11 (total RMSD = 0.63 kcal/mol).
We compare our results against a combinatorially optimized density functional, namely $\omega$B97M-V,\cite{Mardirossian2016Jun} which was found to be the best functional out of 200 functionals examined in Ref.~\citenum{Mardirossian2017}. 
While there are now even more accurate functionals (e.g. double hybrid functionals~\cite{Mardirossian2018Jun,Martin2020Aug}), we believe that $\omega$B97M-V serves as an example of an accurate functional for the types of systems we investigate here. Data for other functionals are available in Ref.~\citenum{Mardirossian2017}. $\omega$B97M-V provides chemical accuracy only for the SN13 subset and within the complete W4-11 set its RMSD is 2.52 kcal/mol.
AFQMC performs quite similarly to $\omega$B97M-V in that
it achieves chemical accuracy for the SN13 subset but not for any other subsets.
Overall, AFQMC's RMSD is 3.26 kcal/mol which is slightly worse than $\omega$B97M-V.
However, we note that this is still an improvement over CCSD (5.40 kcal/mol) by a sizable margin.
As the data points in TAE140 completely determine relative energies in all the other subsets, we see qualitatively similar statistical results between the TAE140 subset and the entire W4-11 data set.

\subsubsection{Multi-reference data points}
\begin{figure}[!h]
    \centering
    \scalebox{0.49}{\includegraphics{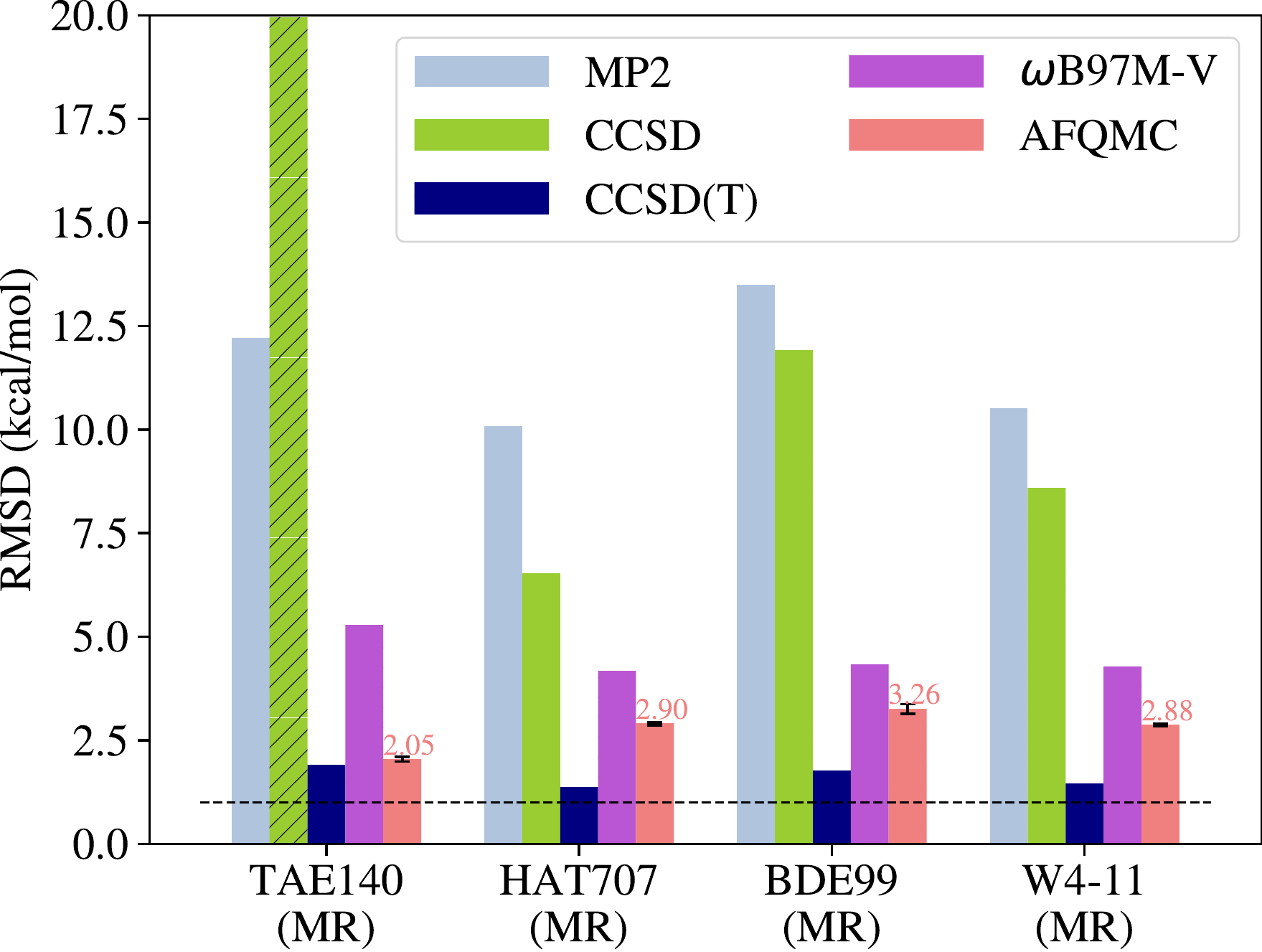}}
    \caption{Root-mean-square deviation (RMSD) (in kcal/mol) forthe MR data points in the TAE140, HAT707, and BDE99 subsets of of the W4-11 data set, along with the overall aggregate W4-11 MR data points. The black dotted lines indicate the ``chemical accuracy'' value of 1 kcal/mol.}
    \label{fig:w411mr}
\end{figure}

For 234 MR data points, we performed the same analysis as shown in~\cref{fig:w411mr}.
With the exception of the error of CCSD for the TAE140 MR set, the relative performance between different methods over MR data points is the same as that over each of the SR data sets and within W4-11 overall. 
Similarly to the SR case shown in~\cref{fig:w411nonmr},
MP2 (10.51 kcal/mol) and CCSD (8.59 kcal/mol) are found to be significantly less accurate than CCSD(T) overall.
CCSD(T) also does not perform as well for these data points, yielding an RMSD of 1.44 kcal/mol. 
$\omega$B97M-V has an RMSD of 4.27 kcal/mol overall for the MR W4-11 data set.
AFQMC is more accurate than all but CCSD(T) for this subset, with an RMSD of 2.88 kcal/mol.
Although AFQMC does not perform as well as CCSD(T) for these data points, it is significantly more accurate than the other approaches considered here.

The good performance of CCSD(T) for these MR data points may be understood by the fact that
many of these MR calculations are not necessarily strongly correlated ones.
There are multiple ways to diagnose MR character, with different metrics providing distinct classifications.
For example, if we use regularized orbital-optimized MP2 ($\kappa$-OOMP2)~\cite{Lee2018Oct} and inspect the underlying spin-symmetry breaking of the solution,
only 7 out of the 16 MR data points in the TAE140 set exhibit spin-symmetry breaking. 
Based on this, one would conclude that only 7 data points in TAE140 should be considered to carry MR character.
We will discuss bond breaking examples in the later sections which will unambiguously fall into the strongly correlated category for stretched geometries.

\begin{figure}[!h]
    \centering
    \scalebox{0.49}{\includegraphics{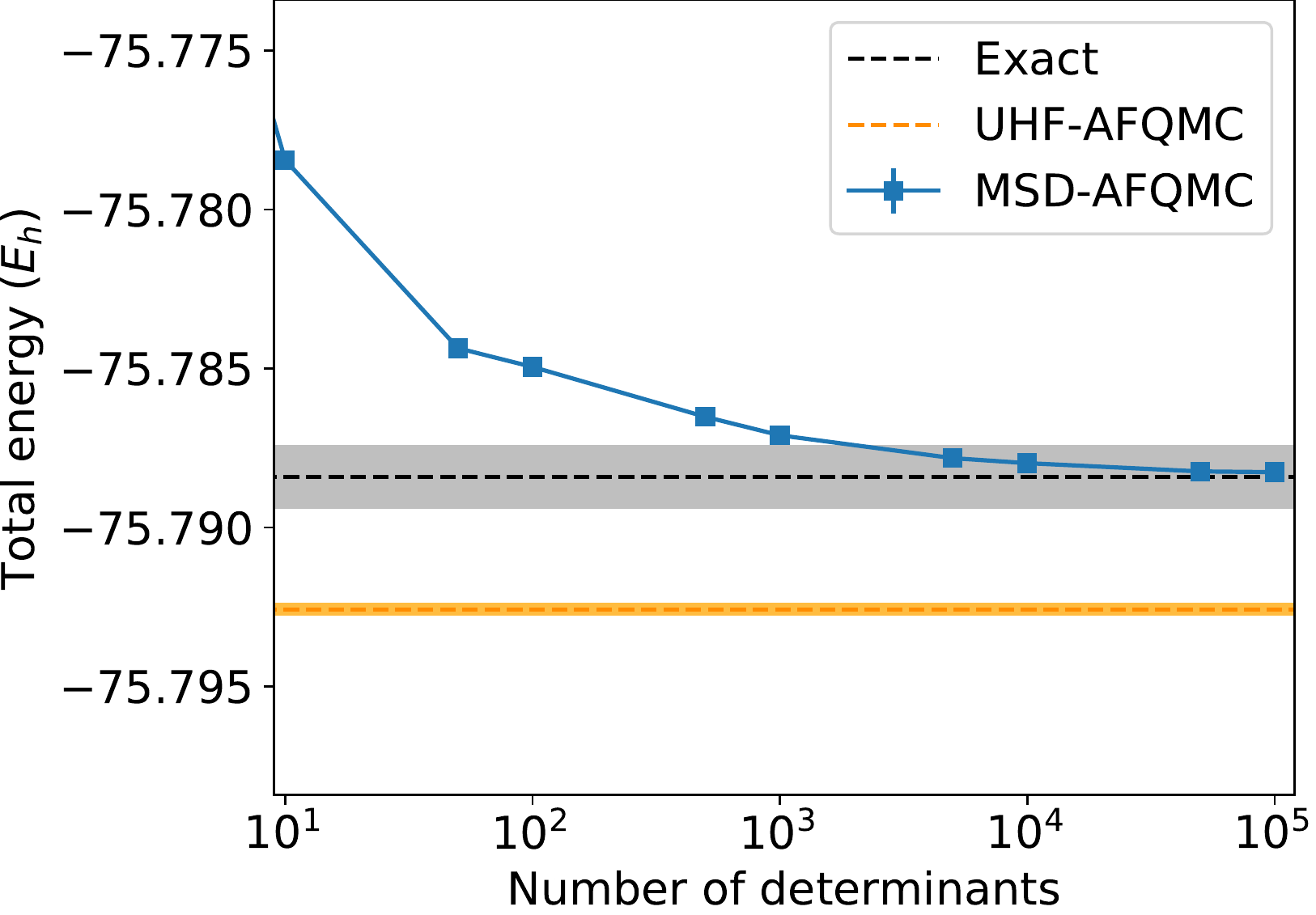}}
    \caption{Convergence of AFQMC energy ($E_h$) with an MSD trial to the exact energy for \ce{C2} in aug-cc-pVTZ as a function of the number of determinants in the trial. Gray area indicates chemical accuracy (1 kcal/mol) from the exact answer. Dotted orange line and orange area denote UHF-AFQMC energy and its error bar. UHF-AFQMC and MSD-AFQMC mean AFQMC performed with UHF and MSD trials, respectively. }
    \label{fig:c2}
\end{figure}

To emphasize the flexibility of AFQMC, we further study \ce{C2} in aug-cc-pVTZ which is one of the representative MR examples in the MR16 subset. We generate a large MSD trial using SHCI for an active space with 8-electron and 90-orbital. We then systematically converge AFQMC energies using that trial to the exact answer obtained by SHCI. As shown in \cref{fig:c2}, MSD-AFQMC quickly improves its accuracy as one adds more determinants to the trial and becomes chemically accurate with $10^4$ determinants or so. Compared to more conventional quantum chemistry methods such as coupled-cluster theory, the generalization of AFQMC to MSD trials is rather straightforward and it can often be used to improve the results significantly beyond AFQMC with an SD trial. One can also try to find other MSD trials generated from a smaller active space to reduce the number of determinants needed to reach chemical accuracy.\cite{Purwanto2009Mar} We will see the power of MSD-AFQMC more later in \cref{subsub:mr}.

\begin{figure}[!h]
    \centering
    \scalebox{0.49}{\includegraphics{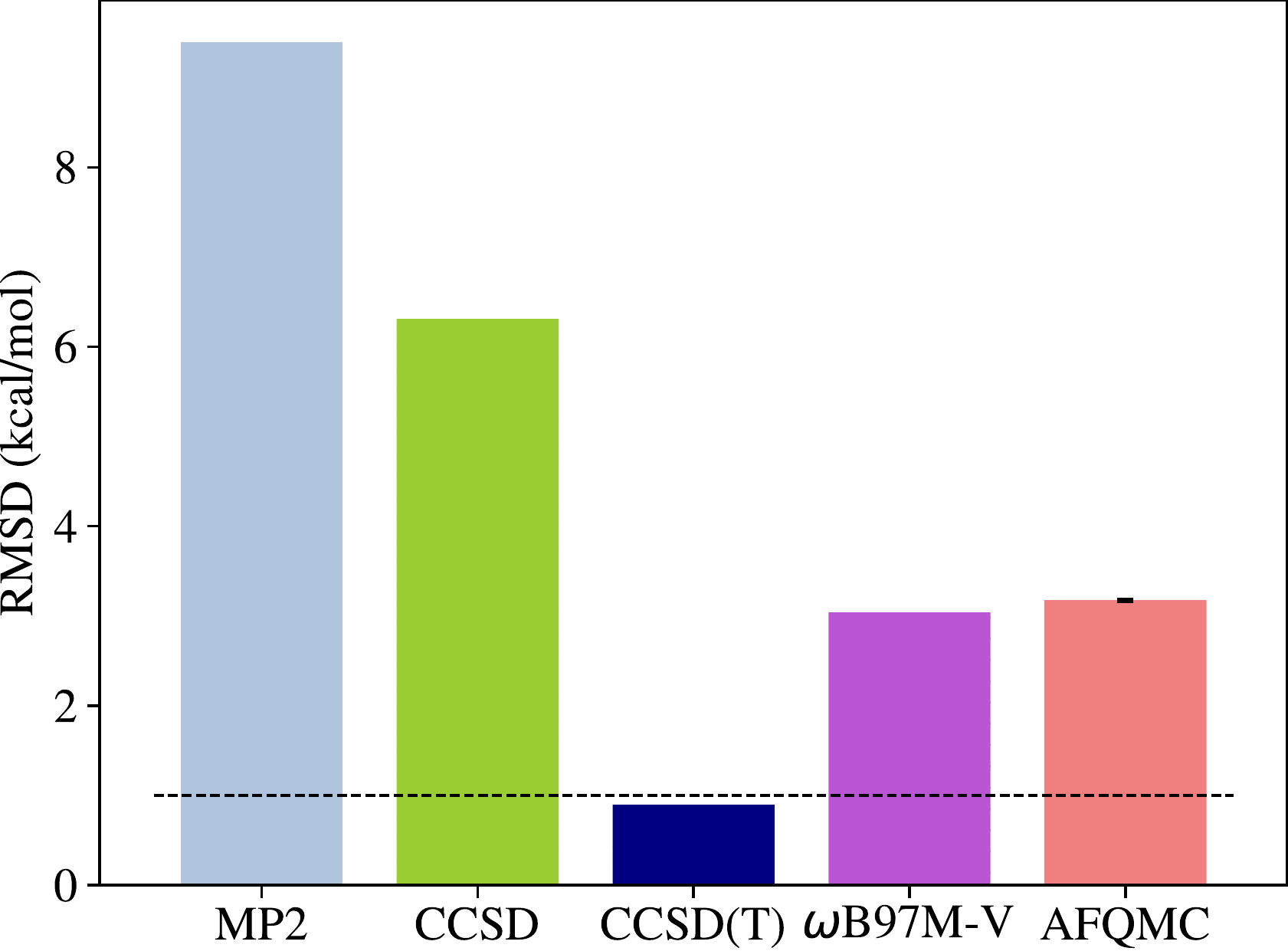}}
    \caption{Root-mean-square deviation (RMSD) for the entire W4-11 benchmark set. The black dotted lines indicate the ``chemical accuracy'' of of 1 kcal/mol.}
    \label{fig:w411}
\end{figure}
In summary, based on this thermochemistry benchmark study,
we conclude that the performance of AFQMC with a simple SD trial is expected to provide accuracy that lies between CCSD and CCSD(T).  It is also quite competitive with a state-of-the-art density functional, $\omega$B97M-V, which is {\em a priori} expected to be accurate for the class of systems studied here.
These conclusions are summarized in~\cref{fig:w411}.
We thus recommend AFQMC with a simple SD trial for calculations where DFT is expected to struggle (either due to self-interaction error or due to strong correlation) and CCSD(T) is too expensive.

\subsection{Non-covalent interaction benchmark (A24)}
\begin{figure}[!h]
    \centering
    \scalebox{0.49}{\includegraphics{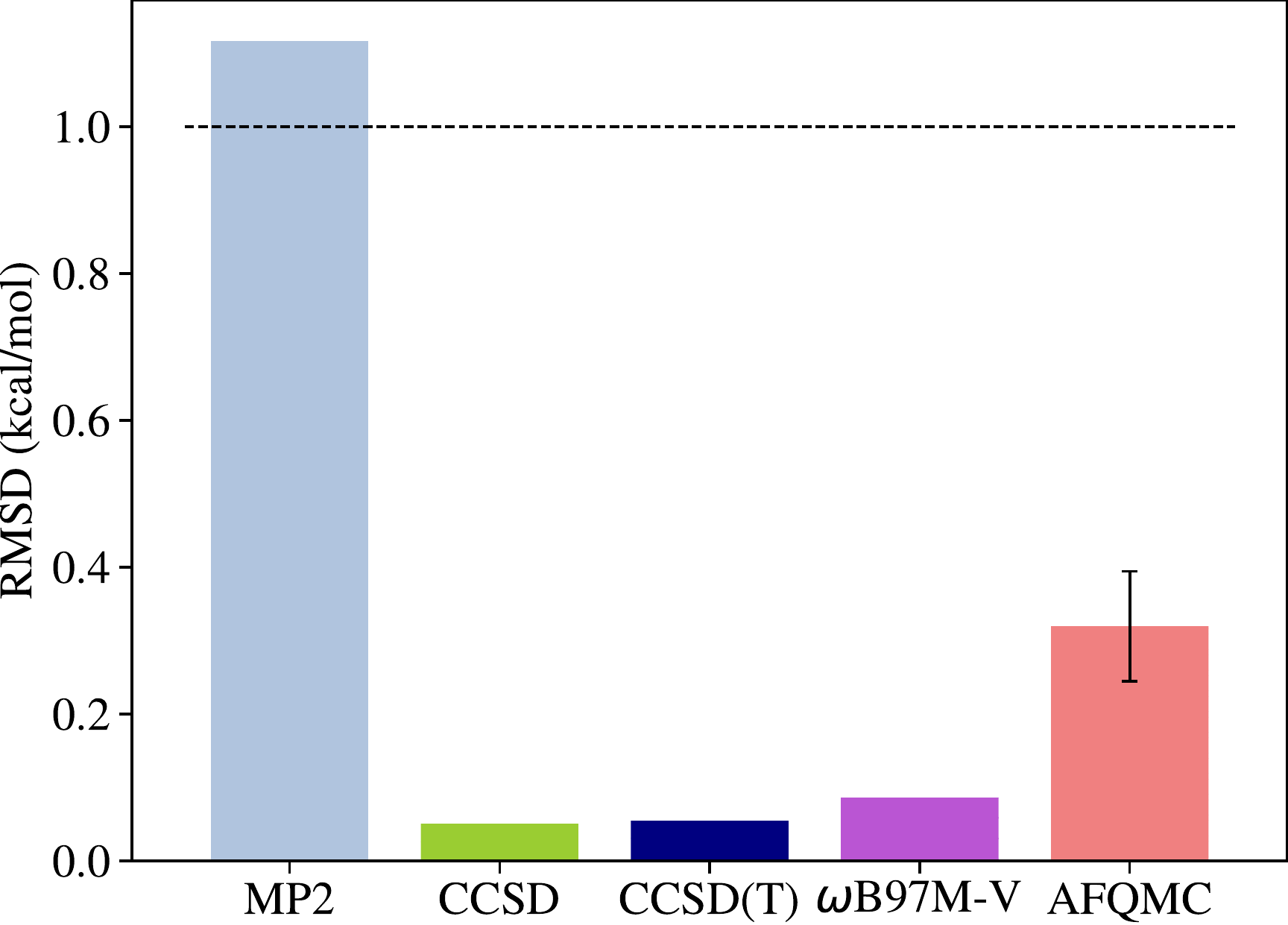}}
    \caption{Root-mean-square deviation (RMSD) for the A24 benchmark set. The black dotted lines indicate the ``chemical accuracy'' value of 1 kcal/mol.}
    \label{fig:a24}
\end{figure}

A24 is a high-quality non-covalent interaction benchmark set composed of 24 small, non-covalently bound complexes.~\cite{rezac2013describing}  All molecules in the set should be well described by RHF wave functions. 
For this set we employ RHF wave functions for both CC and AFQMC methods. We also used the counterpoise correction to eliminate the basis set superposition error.
The energy scale in this set is quite small, with interaction energies ranging from 1.115 kcal/mol to -6.493 kcal/mol.
In~\cref{fig:a24}, we present the RMSD values of MP2, CCSD, CCSD(T), $\omega$B97M-V, and AFQMC for this set.
MP2 has the largest RMSD (1.12 kcal/mol), while CCSD and CCSD(T) are nearly exact (RMSD = 0.05 kcal/mol). 
$\omega$B97M-V is also very accurate with an RMSD of 0.09 kcal/mol. 
AFQMC exhibits more sizable errors than do the CC methods, with an RMSD of 0.32(7) kcal/mol.
This is about a factor of 4 larger error than $\omega$B97M-V,  although still within chemical accuracy.

For relatively simple non-covalently bound systems, combinatorially optimized density functionals like $\omega$B97M-V work well, approaching the accuracy of CCSD(T).~\cite{Mardirossian2016Jun}
Therefore, we recommend the use of AFQMC for non-covalent interaction energies only if the system under study exhibits additional challenges for DFT, such as those arise in cases with open-shell electrons or self-interaction error.  At the same time, the accuracy for AFQMC in such cases in comparison to CCSD(T) is still under-explored. Given the small energy scale and the small number of data points in A24, a more thorough investigation of of non-covalent interactions with AFQMC is highly desirable. The recent report~\cite{Al-Hamdani2021Jun} by Tkatchenko and co-workers of significant deviations between diffusion Monte Carlo and CCSD(T) for large, non-covalently bound complexes thus provides a worthy target for a future investigation performed with AFQMC.

\subsection{Breaking chemical bonds}\label{sub:bondbreaking}
Bond breaking in simple molecular systems has been a test bed for the electronic structure treatment of strong correlation. This is because one can
simply tune the strength of the electron correlation (i.e., the degree to which an SR wave function fails qualitatively) by changing the bond distance.
This fact was recognized early in the development of AFQMC, as evidenced by the work of Zhang and co-workers.~\cite{Al-Saidi2007Oct,Purwanto2008Mar,Purwanto2009Mar,Purwanto2015Feb,Purwanto2016Jun}
In this section, we add more data to this body of work by considering
\ce{H4} in STO-3G and cc-pVQZ bases and \ce{N2} (all-electron) in STO-3G and cc-pVTZ bases.
STO-3G is a minimal basis set and is known to exaggerate strong correlation effects.~\cite{Lee2017Feb}
Therefore, bond breaking in the STO-3G basis set has been a particularly popular test for strong correlation methods.~\cite{Jankowski1980Nov,Siegbahn1983Jun,paldus1993application,Mahapatra1999Apr,kowalski2000complete,Evangelista2006Oct,Ma2006Jan,Small2012Sep,Lee2019Jan}
Larger basis set examples will exhibit
a mixture of strong and weak correlation effects, which is a common challenge
encountered in realistic problems.

The geometry of $\ce{H4}$ in \AA~is
\begin{align*}
\text{H1}&: (0,0,0),\\
\text{H2}&: (0,0,1.23),\\
\text{H3}&: (R_\mathrm{H-H},0,0),\\
\text{H4}&: (R_\mathrm{H-H},0,1.23),
\end{align*}
where we vary the distance between two stretched \ce{H2} units, with each unit with a fixed bond distance of 1.23 \AA.  This value is selected because for $R_\mathrm{H-H} = 1.23$ \AA~, two RHF determinants become quasi-degenerate, maximizing the strong correlation aspect of this model.
Thus RHF is qualitatively incorrect and UHF cannot fully remedy this problem.
A natural active space for this problem is (4e,4o) (i.e., 4-electrons in 4-orbitals),
and this is what we use to generate
complete active space self-consistent field (CASSCF) states for our subsequent 
dynamic correlation treatment. 

In the case of \ce{N2}, we vary the bond distance, $R_\mathrm{N-N}$, between the two nitrogen atoms.
As the bond is stretched, \ce{N2} exhibits strong correlation effects, and ultimately dissociates to two quartet nitrogen atoms. 
RHF cannot correctly dissociate \ce{N2} whereas UHF dissociates this molecule correctly.
For large bond distances, UHF is therefore expected to be qualitatively correct, while RHF qualitatively fails. Nonetheless, for intermediate bond distances, UHF is incapable of describing spin-recoupling and will thus lead to quantitatively inaccurate results in this regime. For CASSCF, we use a minimal active space of (6e,6o) (i.e., 6-electrons in 6-orbitals).

\begin{table*}[t]
\begin{tabular}{|c|c|c|}
\hline
Acronym & Full Name & References\\\hline
CASPT2 & Complete Active Space Second-Order Perturbation Theory & \citenum{Andersson1990Jul,Andersson1992Jan}\\ \hline
NEVPT2 & N-electron Valence Second-Order Perturbation Theory & \citenum{Angeli2001Jun,Angeli2002Nov}\\ \hline
MRMP2 & \begin{tabular}[c]{@{}c@{}}Multireference Second-Order  \\ M{\o}ller-Plesset Perturbation Theory\end{tabular} & \citenum{Hirao1992Mar}\\ \hline
MRACPF& Multireference Average Coupled Pair Functional & \citenum{Gdanitz1988Jan} \\ \hline
MRAQCC& Multireference Average Quadratic Coupled Cluster & \citenum{Szalay1993Nov,Szalay1995Sep} \\ \hline
MRCISD+Q & \begin{tabular}[c]{@{}c@{}}Multireference Configuration Interaction \\ with Singles, Doubles, and   Davidson correction\end{tabular} & 
 \citenum{Bruna1980Jun,Burton1983Mar}\\\hline
DSRG-MRPT2 & \begin{tabular}[c]{@{}c@{}}Driven Similarity Renormalization Group \\ Multiereference Second-Order Perturbation Theory\end{tabular} & \citenum{Li2015May} \\\hline
DSRG-MRPT3 & \begin{tabular}[c]{@{}c@{}}Driven Similarity Renormalization Group \\ Multiereference Third-Order Perturbation Theory\end{tabular} &
\citenum{Li2017Mar}\\\hline
MR-LDSRG(2) & \begin{tabular}[c]{@{}c@{}} Multiereference Linearized Driven Similarity \\ Renormalization Group Truncated with Two-Body Operators\end{tabular} & \citenum{Li2016Apr}\\\hline
\end{tabular}
\caption{
Summary of multi-reference methods considered for bond-breaking benchmark problems in this work.
}
\label{tab:methods}
\end{table*}
To add dynamic correlation on top of the CASSCF calculation, we consider some of the more widely used multi-reference methods in addition to AFQMC.
These methods are summarized in \cref{tab:methods} along with their acronyms.
For CASPT2 we did not use any shifts.~\cite{Roos1995Oct,Roos1996Dec,Forsberg1997Aug,Ghigo2004Sep}
NEVPT2 is performed via the partially contracted algorithm,~\cite{Angeli2002Nov} whereas the rest of the deterministic MR methods employed fully internally contracted versions.
These methods represent commonly used dynamic correlation approaches for MR systems.

\subsubsection{Assessment of single-reference methods}

\begin{figure}[!h]
    \centering
    \scalebox{0.49}{\includegraphics{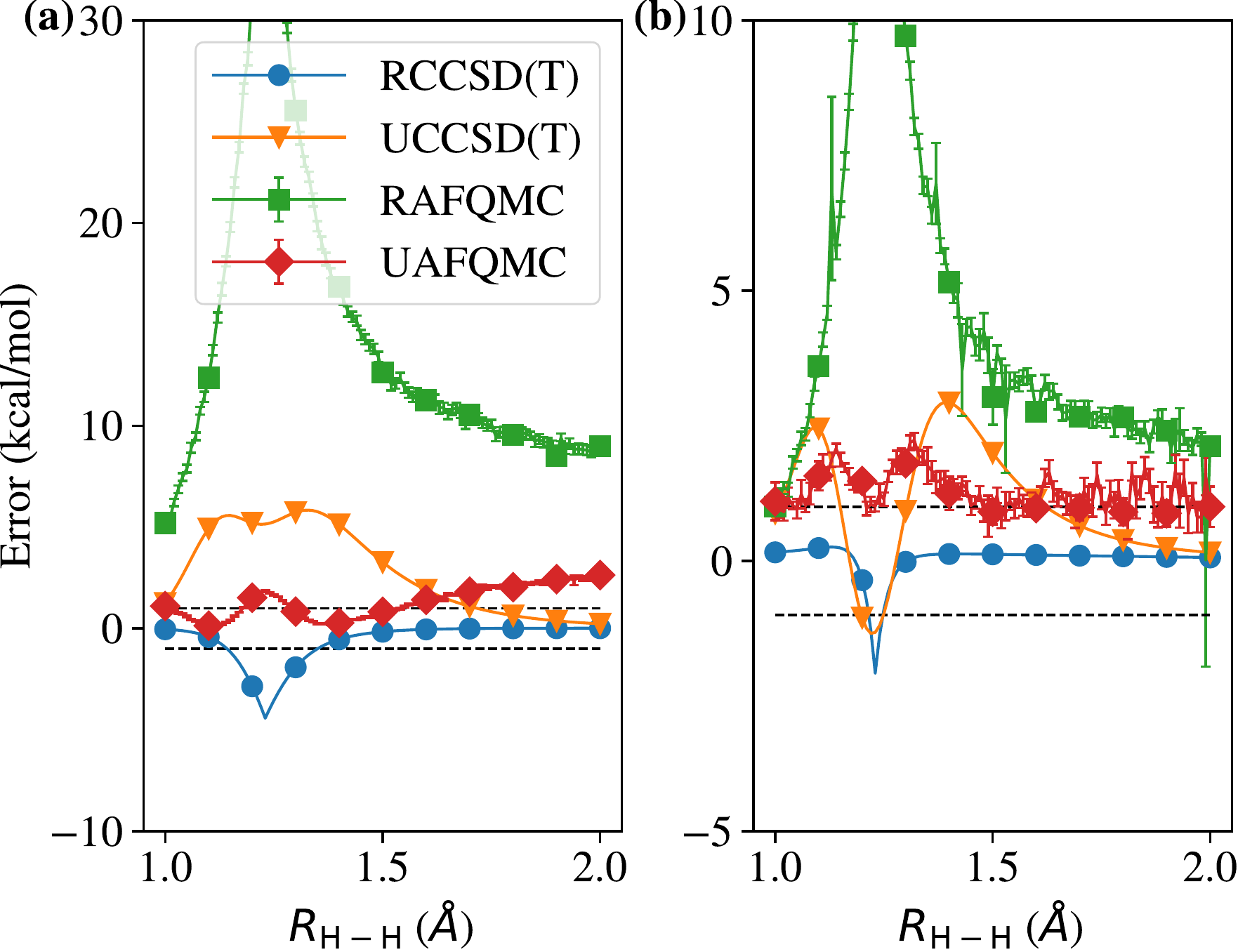}}
    \caption{
    Error in the potential energy curve of H$_4$ as a function of $R_\mathrm{H-H}$ for methods based on HF states (a) using a STO-3G basis set and (b) using a cc-pVQZ basis set.
    Note that there are more data points than markers shown.
    See \cref{fig:pesabs} for qualitative shapes of these potential energy curves in absolute energy scale.
    }
    \label{fig:h4sr}
\end{figure}

We first discuss the performance of single-reference methods for \ce{H4}.
In~\cref{fig:h4sr} (a), we present the error for each single-reference method for the STO-3G basis set. We examine both RHF and UHF states for CCSD(T) and AFQMC for use as reference or trial functions.
These different calculations are referred to as RCCSD(T)/UCCSD(T) and RAFQMC/UAFQMC, respectively.

The sharp derivative discontinuity at $R_\mathrm{H-H} =$ 1.23 \AA~ in RCCSD(T) and RAFQMC is due to the two RHF solutions crossing at $R_\mathrm{H-H} =$ 1.23 \AA.
As noted previously in literature,~\cite{Small2012Sep} RCCSD(T) becomes non-variational as the unit bond distance approaches $R_\mathrm{H-H} =$ 1.23 \AA. 
In the case of RAFQMC, the energy is always above the exact energy, but the magnitude of the error is about nine to ten times larger than that of RCCSD(T). 
UCCSD(T) produces variational energies at every distance considered here.
UAFQMC performs far better than RAFQMC at every bond distance and also
outperforms UCCSD(T) up to $R_\mathrm{H-H} =$ 1.64 \AA.  This is highly encouraging because UAFQMC is very accurate near the distances where strong correlation is most pronounced.
Nonetheless, UAFQMC does not dissociate this molecule into two independent \ce{H2} correctly.
Both RCCSD(T) and UCCSD(T) correctly describe dissociation, but neither does RAFQMC nor UAFQMC.
In~\cref{subsub:atom}, we discussed
the exactness of CCSD(T) and the non-exactness of AFQMC
for two-electron systems.
The \ce{H4} $\rightarrow$ 2 \ce{H2} dissociation problem is thus another good illustration of this point.

We observe similar behavior of all four methods in the larger basis set, cc-pVQZ, as illustrated in~\cref{fig:h4sr} (b).
Since strong correlation is less pronounced in a larger basis set, 
the magnitude of the error produced by all methods is significantly smaller compared to that exhibited  in the STO-3G basis.
Nonetheless, qualitative failures of RCCSD(T), such as non-variationality and a derivative discontinuity at the square geometry are still observed in this basis set.
As is the case for the STO-3G basis, RAFQMC exhibits a significantly larger error than RCCSD(T) does in this basis set.  Finally, we observe quite similar levels of quantitative accuracy when comparing UCCSD(T) to UAFQMC.

\begin{figure}[!h]
    \centering
    \scalebox{0.48}{\includegraphics{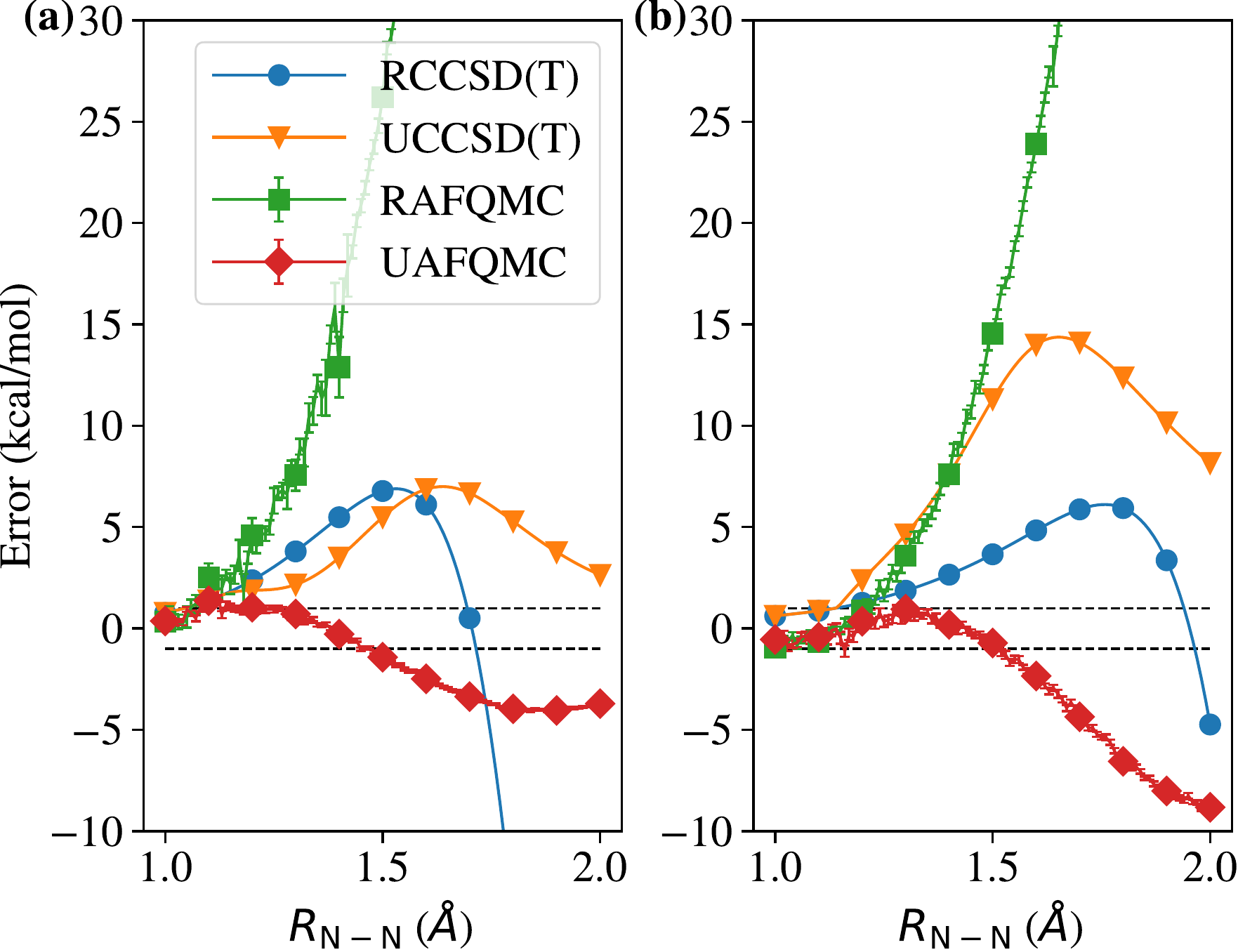}}
    \caption{Error in the potential energy curve of N$_2$ as a function of $R_\mathrm{N-N}$ for methods based on HF states (a) using an STO-3G basis set and (b) using a cc-pVTZ basis set.
        Note that there are more data points than markers shown.
    See \cref{fig:pesabs} for qualitative shapes of these potential energy curves in absolute energy scale.
        }
    \label{fig:n2sr}
\end{figure}

We continue the discussion of the performance of single-reference methods for \ce{N2} as shown in ~\cref{fig:n2sr} for (a) STO-3G and (b) cc-pVTZ.
Unlike \ce{H4}, we find that the errors exhibited by different methods are comparable between these two bases sets.
While there are multiple RHF solutions for \ce{N2}, we focused on the one that does not break spatial symmetry.
RCCSD(T) exhibits incorrect turnover behavior and non-variationality at far distances, whereas UCCSD(T) has a large energy error in the spin-recoupling regime.
UCCSD(T) ultimately dissociates correctly into two N atoms, but its error up to a bond distance $R_\mathrm{N-N}$ = 2 \AA~ is quite sizable (7.5 kcal/mol for STO-3G and 15 kcal/mol for cc-pVTZ, respectively).  RAFQMC again performs significantly worse than RCCSD(T).
UAFQMC is quantitatively comparable to UCCSD(T) for both bases sets, but the sign of its error is opposite to that of UCCSD(T).  We note that the fact that UAFQMC exhibits non-variationality is not unexpected, as AFQMC energies are not guaranteed to be variational.~\cite{carlson_no_var}

\begin{table}[h!]
\begin{tabular}{|l|rr|rr|}
\hline
\multicolumn{1}{|c|}{} & \multicolumn{2}{c|}{\ce{H4}}                                    & \multicolumn{2}{c|}{\ce{N2}}                                   \\ \cline{2-5} 
                       & \multicolumn{1}{c|}{STO-3G} & \multicolumn{1}{c|}{cc-pVQZ} & \multicolumn{1}{c|}{STO-3G} & \multicolumn{1}{c|}{cc-pVTZ} \\ \hline
RCCSD(T)               & \multicolumn{1}{r|}{4.43}   & 2.33                         & \multicolumn{1}{r|}{74.95}  & 10.96                        \\ \hline
UCCSD(T)               & \multicolumn{1}{r|}{5.61}   & 4.28                         & \multicolumn{1}{r|}{6.23}   & 13.80                        \\ \hline
RAFQMC                 & \multicolumn{1}{r|}{35.37(5)}  & 18.65(36)                        & \multicolumn{1}{r|}{140.20(14)} & 87.72(23)                       \\ \hline
UAFQMC                 & \multicolumn{1}{r|}{2.51(8)}   & 1.57(30)                         & \multicolumn{1}{r|}{5.89(16)}   & 9.79(22)                         \\ \hline
\end{tabular}
\caption{
Non-parallelity error (maximum error - minimum error) in kcal/mol for single-reference methods.
}
\label{tab:srnpe}
\end{table}

In summary, for bond breaking examples and other typical strongly correlated systems, 
it appears that RAFQMC exhibits significant errors, much larger than what is seen with RCCSD(T).
However, UAFQMC is quantitatively better than UCCSD(T), although it is still far from chemical accuracy.  This can be seen from the non-parallelity errors presented in~\cref{tab:srnpe}.
Non-parallelity error is a commonly used metric for assessing the performance of different methods in the computation of the relative energetics on a
potential energy surface.
Therefore among single-reference methods, UAFQMC is a promising candidate for the treatment of relatively simple strongly correlated systems.

\subsubsection{Assessment of multi-reference methods}\label{subsub:mr}

\begin{figure}[h!]
    \centering
    \scalebox{0.49}{\includegraphics{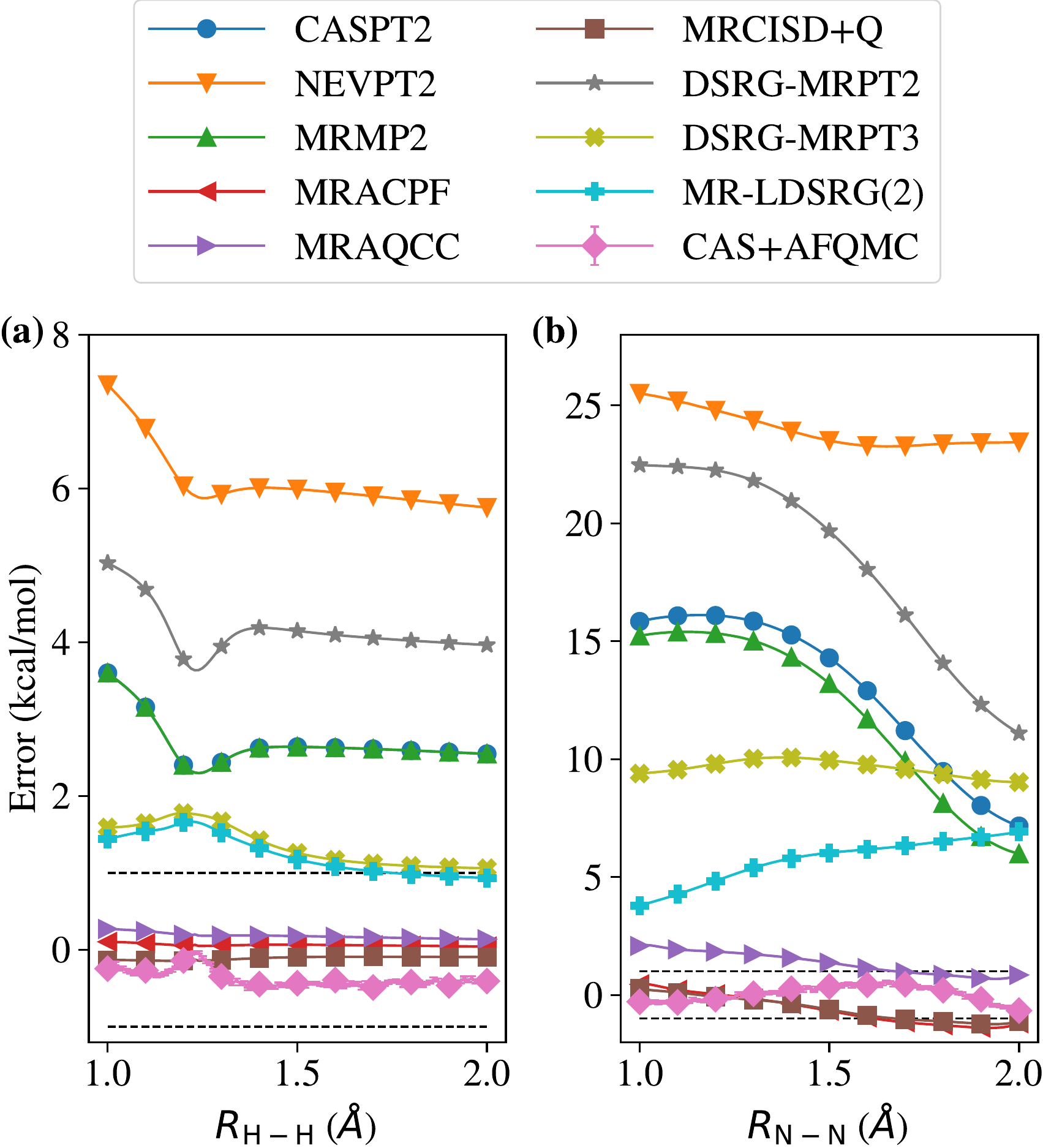}}
    \caption{Error in the potential energy curve of methods based on CASSCF states (a) \ce{H4} using a cc-pVQZ basis set and (b) \ce{N2} using a cc-pVTZ basis set.
    For \ce{H4}, MRMP2 and CASPT2 yield identical data.
    Note that there are more data points than markers shown.
        See \cref{fig:pesabsmr} for qualitative shapes of these potential energy curves.
    }
    \label{fig:bondmr}
\end{figure}

We repeat the same analysis for MR methods where simple CASSCF states are employed either as a reference state or a trial wave function, and quantify the errors associated with the treatment of dynamic correlation.
In~\cref{fig:bondmr}, it is clear that the worst performer in terms of absolute energies in both \ce{H2} and \ce{N2}
is NEVPT2.  CASPT2 and MRMP2 perform much better than NEVPT2,  by approximately 3 kcal/mol for \ce{H2} and by more than 10 kcal/mol for \ce{N2}.
Nonetheless, in terms of non-parallelity error (NPE), see \cref{tab:mrnpe},
NEVPT2 is comparable to CASPT2 and MRMP2 for \ce{H4} and
is better than these two methods for \ce{N2} by more than 6 kcal/mol.
CASPT2 and MRMP2 methods perform quite similarly (identically for \ce{H4}), and their performance is far from the level of chemical accuracy.  The performance of DSRG-MRPT2 falls between that of NEVPT2 and CASPT2/MRPT2.
The poor performance of second-order perturbation theories is particularly worrying,
as these methods are the most commonly employed MR methods due to their simplicity and their relatively inexpensive nature, namely a scaling of $\mathcal O(N^5)$ outside the active space.

We have also investigated the performance of MRCISD using different ways to correct for its size-inconsistency error (namely MRACPF, MRAQCC, MRCISD+Q).
These methods are computationally more expensive than MRPT methods, with a scaling of $\mathcal O(N^6)$ outside the active space.
This scaling also applies to two DSRG methods, DSRG-MRPT3 and MR-LDSRG(2).
MRACPF, MRAQCC, and MRCISD+Q are all nearly exact for \ce{H4}.
MRAQCC becomes more inaccurate for \ce{N2}, exhibiting errors beyond chemical accuracy at small bond distances.
Nonetheless, MRACPF and MRCISD+Q provide near-chemical acuracy for all bond distances for \ce{N2}.
For both molecules, MR-LDSRG(2) is not as accurate as the MRCI methods for both absolute and relative energies.
Nonetheless, for \ce{H4}, its NPE is only 0.73 kcal/mol.
DSRG-MRPT3 performs very similarly to MR-LDSRG(2) for \ce{H4}, but it differs significantly from MR-LDSRG(2) for \ce{N2}
For \ce{N2}, DSRG-MRPT3 produces absolute energies that are quite far from chemical accuracy, but its NPE is quite small and comparable to that of MRCISD+Q and MRAQCC.

AFQMC with CASSCF trial wave functions 
works remarkably well for both \ce{H2} and \ce{N2}
achieving chemical accuracy at all bond distances.
This accuracy comes with a far lower scaling cost (see \cref{tab:cost}) than both MRPT and MRCI-based methods, where we emphasize the steep scaling of these methods within the active space due to the requirement of higher-order density matrices.
We also note that none of the other MR methods except NEVPT2 and the DSRG methods are size-consistent.
Since CASSCF is size-consistent, AFQMC with CASSCF trial wave functions is also size-consistent in the limit of $\Delta t \rightarrow 0$ (see \cref{subsec:size}).

\begin{table}[]
\begin{tabular}{|c|r|r|}
\hline
          & \multicolumn{1}{c|}{\ce{H4}} & \multicolumn{1}{c|}{\ce{N2}} \\ \hline
CASPT2    & 1.30                    & 8.98                    \\ \hline
NEVPT2    & 1.59                    & 2.29                    \\ \hline
MRMP2     & 1.30                    & 9.43                    \\ \hline

DSRG-MRPT2     &      1.39               & 11.38                    \\ \hline
DSRG-MRPT3    &        0.72             & 1.16                    \\ \hline
MR-LDSRG(2)    &        0.73             & 3.16                    \\ \hline

MRACPF    & 0.06                    & 2.08                    \\ \hline
MRAQCC    & 0.13                    & 1.62                    \\ \hline
MRCISD+Q  & 0.05                    & 1.67                    \\ \hline
CAS+AFQMC & 0.48(2)                    & 1.27(8)                    \\ \hline
\end{tabular}
\caption{
Non-parallelity error (maximum error - minimum error) in kcal/mol for multi-reference methods.
}
\label{tab:mrnpe}
\end{table}

Due to its computational efficiency and its accuracy in reproducing both absolute energies and NPE values (see~\cref{tab:mrnpe}), 
AFQMC would appear to be the method of choice when strong correlation can be handled by the CASSCF trial wave function and the remaining dynamic correlation needs to be corrected at scale. With the improved algorithm for using SCI trial wave functions,~\cite{Mahajan2022May}
AFQMC will likely become a commonly used MR dynamic correlation method in the future.

\section{Outlook and future opportunities}\label{sec:lessons}
There are many future opportunities for developments and improvements of different aspects of ph-AFQMC.
Due to its relatively new vintage and the comparatively limited use of the method in quantum chemistry, the development of ph-AFQMC when compared to, for example, CC methods has not yet reached full maturity.
We note several promising opportunities for future development in this section.

\begin{enumerate}
\item {\it Trial wave functions:}~
As mentioned in~\cref{subsec:trial},
the form of trial wave functions is currently limited
to SD trials or relatively compact MSD trials. 
Using more accurate trial wave functions without greatly increasing the overall computational cost is a forefront area for future ph-AFQMC development.
Compared to CCSD(T), where going beyond the SD reference is formally challenging, the flexibility afforded by ph-AFQMC will lead to important developments along this line. Our experience suggests that the sensitivity to the trial wave function within ph-AFQMC is far greater than is the sensitivity of CC methods to the reference wave functiton.  This both provides opportunities to greatly enhance accuracy but also challenges for rationally developing strategies for the use of trial functions. 
\item {\it Alternative constraints:}~
Due to the lack of extensive benchmark studies, we do not know if the constraint provided by the cosine projection in~\cref{eq:mod_overlap}
is optimal in terms of accuracy and statistical efficiency. As noted in~\cref{subsec:size}, the currently used phaseless constraint introduces an $\mathcal O(\Delta t)$ time step error, limiting the size of time steps and necessitating time step extrapolations beyond the Trotter error. Furthermore it may well be that different constraints provide different levels of accuracy even when used with the same trial wave function.  Much more work in this direction is warranted.
\item {\it Observables other than ground state energies:}~ Some observable properties extracted from ph-AFQMC have been shown to be less reliable than is desirable.~\cite{Motta2017Nov,Lee2021Jun}
The accuracy of properties computed via back propagation provides one concrete example. 
Furthermore, excited state energies also fall into this category where additional complications arise due to the nature of projector Monte Carlo.
At this stage ph-AFQMC is an accurate method for the ground state energy in chemical systems, however ancillary algorithms that enable the computation of other properties with the accuracy consistent with that of the ground state energy within ph-AFQMC is an important goal for future research.

\item {\it Computational cost:}~
Local energy evaluation algorithms for ph-AFQMC with SD trial wave functions
are relatively mature and provide multiple ways to reach cubic scaling per sample.~\cite{Motta2019Jun,Malone2019Jan,Lee2020Jul,Weber2022Jun}
However, algorithms to accelerate walker propagation are scarce.  In addition, 
local energy evaluation algorithms for ph-AFQMC with more complex MSD trial wave functions have been relatively less explored.~\cite{Weber2022Jun,Mahajan2022May}
Thus, reducing the computational cost of all components in ph-AFQMC other than the local energy evaluation forms another direction worthy of future effort.
\end{enumerate}

\section{Conclusions}\label{sec:conclusions}
The purpose of this work has been two-fold: First, we have presented a self-contained overview of the formalism behind the ph-AFQMC approach from a quantum chemistry perspective, and have delineated the considerations needed to understand the computational implementation and cost of the approach.  Second, we have assessed the performance of ph-AFQMC for a well-known main group chemistry benchmark set (W4-11) and a non-covalent interaction benchmark set (A24), leading to a total of 1004 relative energies.
In addition, we have studied the potential energy curves of 
commonly studied model problems, namely \ce{H4} and \ce{N2}, documenting the performance of ph-AFQMC with different trial wave functions
against other standard wave function-based quantum chemistry methods.  This constitutes the largest quantum chemical benchmark study to date using ph-AFQMC. While
the knowledge gained from our work is far from complete,
we will conclude our work with some cautious recommendations for the use of ph-AFQMC in a broad chemical context.

{\it Single Slater determinant trial wave functions:}~
ph-AFQMC employed with a with single determinant trial is an accurate method that can be a method of choice if CCSD(T) is too expensive and if the target system does not exhibit open-shell electrons with sizable antiferromagnetic coupling. 
While studies on chemical systems using DMC and GFMC have been limited, AFQMC was shown to be more accurate than these for the same single determinant trial for several specific systems.\cite{williams2020direct,Malone2020Oct}
Such trials will generally take the form of HF wave functions, or DFT~\cite{Zhang2003Apr,al2006auxiliary} or regularized OOMP2~\cite{Lee2018Oct,Lee2020May} wave functions when HF exhibits unphysical properties such as over-localization.
The performance of ph-AFQMC for non-covalent interaction energies is still
unclear due to the limited number of data points we have considered. We hope that the community will take up the task of a more detailed investigation in the near future.

{\it Multi-Slater determinant trial wave functions:}~
For systems exhibiting open-shell electrons with antiferromagnetic coupling
either due to bond breaking or arising from
localized d- or f-electrons,
ph-AFQMC must be used with more elaborate multi-Slater trial wave functions such as CASSCF or SCI wave functions.
Using spin-unrestricted single determinant trials can be an option if there are only two electrons contributing to the strong correlation behavior.~\cite{Lee2019Jun,Lee2020May}
However, the phaseless error needs to be carefully quantified by more sophisticated trial wave functions in realistic strongly correlated if possible.
We stress that ph-AFQMC was found to be far more accurate than other MRPT methods, such as CASPT2, NEVPT2, and MRMP2.
Given its high efficiency compared to MRCI methods,
we anticipate ph-AFQMC to become a clear method of choice for
including dynamic correlation in active-space calculations.
With improved algorithms~\cite{Mahajan2022May} and implementations leveraging graphical processing units (GPUs),~\cite{Shee2018Aug,Malone2020Jul} we believe that ph-AFQMC with $\sim 10^5$ to $10^6$ determinants will become routine calculations.
Nonetheless, enabling the efficient use of sophisticated trial wave functions is one of the most pressing  topics for near-term development of ph-AFQMC.~\cite{hlubina1997ferromagnetism,chang2016auxiliary,Mahajan2021Aug,Huggins2022Mar}

We hope that the numerical data presented in this work, as well as the existence of currently available open-source packages~\cite{Kent2020May,ipie,Malone2022Sep} will
encourage many more developers and users to explore uncharted territory and contribute to method development within the ph-AFQMC framework.
While the data provided here constitutes the most extensive data set for ph-AFQMC to date, it is important to continue producing data and comparisons for a wide class of systems. 
Furthermore, compared to other deterministic quantum chemistry methods, ph-AFQMC is not yet at the stage of development where it can be considered a  ``black-box'' approach. In addition, its final energy is also statistical in nature, introducing additional barrier
for users to use the method. It is, therefore, important to have and further develop open-source frameworks\cite{ipie,Malone2022Sep}
and to provide straightforward user interfaces.
Such efforts will enable the expansion of our understanding of the relative benefits and weaknesses of ph-AFQMC, and we hope this goal is taken up by the electronic structure community with enthusiasm in the coming years.

\section{Acknowledgment}
We thank Luke Bertels, Garnet Chan, Francesco Evangelista, Fionn Malone, Adam Rettig, Sandeep Sharma, and Shiwei Zhang for helpful discussions. DRR would like to thank Richard Friesner and Shiwei Zhang for collaboration on AFQMC-related topics over the last several years. We thank Richard Friesner for the suggestion to carefully consider basis set extrapolations in this work.  We thank Google for their gift that was used to support JL and HQP in developing a code used in this work.
We also thank Francesco Evangelista for providing the bond dissociation data of \ce{H4} and \ce{N2} for the DSRG methods.
We acknowledge computing resources from Columbia University's Shared Research Computing Facility project, which is supported by NIH Research Facility Improvement Grant 1G20RR030893-01, and associated funds from the New York State Empire State Development, Division of Science Technology and Innovation (NYSTAR) Contract C090171, both awarded April 15, 2010.

\section{Appendix}
\appendix
\beginsupplement

\setcounter{equation}{0}
\renewcommand{\theequation}{A\arabic{equation}}
\setcounter{subsection}{0}
\renewcommand{\thesubsection}{A\arabic{subsection}}

\subsection{Further details on ground-state calculations}\label{appsec:ground}

Discretizing imaginary time, one may write the imaginary time propagation in \cref{eq:imag_proj} as a product of infinitesimal propagators
\begin{equation}
e^{-\tau\hat{H}} = \left(e^{-\Delta \tau\hat{H}} \right)^N,
\label{imag_time_proj}
\end{equation}
where $N$ is the total number of imaginary time steps and the small imaginary time step is defined as $\Delta \tau = \tau /N$. 

The efficiency of AFQMC relies on the representation of the short-time propagator.
One first recasts $\hat{H}$ in \cref{eq:Habinitio} into
\begin{equation}
\hat{H} = \hat{H}_1 - \frac{1}{2} \sum_{\gamma=1}^{N_{\gamma}} \hat{v}_{\gamma}^2,
\label{ham}
\end{equation}
where $\hat{H}_1$ is the one-body part of the Hamiltonian, 
\begin{equation}
\hat{H}_1 = 
\sum_{pq}(h_{pq}-\frac12\sum_r (pr|rq)) a_p^\dagger a_q,
\end{equation}
and the two-body part is written as the sum of $N_{\gamma}$ squared operators, for example, by using the modified Cholesky decomposition~\cite{Beebe1977,aquilante2010molcas} or density fitting,~\cite{Dunlap1979,Weigend2002Feb}
\begin{equation}
\hat{v}_\gamma = \sum_{pr}
L_{pr}^\gamma a_p^\dagger a_r,
\end{equation}
with 
\begin{equation}
(pr|qs) = \sum_{\gamma=1}^{N_\gamma} L_{pr}^\gamma L_{qs}^\gamma.
\end{equation}
In practice,
we modify this Hamiltonian further
by performing a
mean-field subtraction~\cite{Al-Saidi2006Jun} or using a
shifted contour~\cite{Rom1998Nov}.
This amounts to writing
\begin{equation}
\hat{v}_\gamma' =  \hat{v}_\gamma - \langle\Psi_T|\hat{v}_\gamma|\Psi_T\rangle.
\end{equation}

To realize \cref{eq:imag_proj} in a computationally efficient manner, 
we write the global wave function as a weighted summation over a statistical sample of wave functions, $\{|\psi_i\rangle\}$, commonly referred to as ``walkers'':
\begin{equation}
|\Psi(\tau)\rangle
=
\sum_{i=1}^{N_\text{walkers}}
w_i (\tau) 
|\psi_i(\tau)\rangle,
\end{equation}
where $w_i(\tau)$ is the weight of the $i$-th walker at imaginary time $\tau$.
We also express $\hat{B}(\textbf{x})$ in~\cref{eq:Bx} as
\begin{equation}
\hat{B}(\textbf{x}) = \text{exp} \left( -\Delta\tau\hat{H}_1\right) \exp\left( \sqrt{\Delta\tau} \sum_{\gamma=1}^{N_{\gamma}} x_{\gamma} \hat{v}_{\gamma}' \right).
\label{app:Bx}
\end{equation}
An instance of the short-time propagator in \cref{app:Bx} is then applied to a set of random walkers $\{|\psi_i\rangle\}$ to obtain the ground state $|\Psi_{0}\rangle$ from the initial state. 
Namely, each walker is assigned to a vector of Gaussian random variables $\mathbf x_i (\tau)$ at imaginary time $\tau$
and one applies \cref{app:Bx} to advance the wave function to the next time step
\begin{equation}
|\psi_i(\tau+\Delta\tau)\rangle
=
\hat{B}(\textbf{x}_i(\tau))
|\psi_i(\tau)\rangle.
\end{equation}
This procedure effectively computes the high-dimensional integral in \cref{eq:Bx} by means of Monte Carlo sampling.
Single Slater determinants are the most common form of walker wave functions, and the Thouless theorem ensures that the walkers will stay in the single Slater determinant manifold upon the action of the propagator.~\cite{thouless_theorem,thouless_theorem_2}

Global estimates for operators that commute with $\hat{H}$ are evaluated by the following mixed estimator
\begin{equation}
\langle O(\tau)\rangle_\text{mixed} = 
\frac
{
\langle \Psi_T | \hat{O} | \Psi(\tau)\rangle
}
{\langle \Psi_T | \Psi(\tau)\rangle},
\label{eq:Em}
\end{equation}
where $|\Psi_T\rangle$ is a wave function that allows for a straightforward evaluation of the above expression.
If $\hat{O} = \hat{H}$, this provides an unbiased way to estimate the total energy.
The algorithm discussed so far is defined as the free-projection AFQMC (fp-AFQMC) which is exact in principle, however, suffers from the phase (sign) problem.~\cite{Zhang2003Apr} 
Due to this problem, the mixed estimate in \cref{eq:Em} will generically have a variance that scales exponentially with system size.~\cite{Troyer2005May}

One way to control the severe statistical fluctuation arising from the phase problem is to use the phaseless approximation,\cite{Zhang2003Apr}  resulting in the phaseless AFQMC method (ph-AFQMC). 
ph-AFQMC utilizes importance sampling based on a trial wave function $|\Psi_T\rangle$ throughout the random walk process
and thereby works with the statistical representation of global wave functions as in \cref{eq:stat_wfs}.
Given this importance sampling,
the walker weight update rule follows
\begin{equation}
w_{i}(\tau+\Delta\tau) = 
w_i(\tau) \times
I ( \textbf{x}_i(\tau), \bar{\textbf{x}}_i(\tau), |\psi_{i}{(\tau)}\rangle ),
\label{import_sampling}
\end{equation}
where the importance function $I$ which is proportional to the overlap ratio in~\cref{eq:overlap} is given by
\begin{align}\nonumber
I ( \textbf{x}_i(\tau), \bar{\textbf{x}}_i(\tau), |\psi_{i}{(\tau)}\rangle ) &=
S ( \textbf{x}_i(\tau)- \bar{\textbf{x}}_i(\tau), |\psi_{i}{(\tau)}\rangle )\\
&\times e^{\textbf{x}_i(\tau) \bar{\textbf{x}}_i(\tau) - \bar{\textbf{x}}_i(\tau) \bar{\textbf{x}}_i(\tau) /2}.
\label{app:import_func}
\end{align}
$\bar{\textbf{x}}_i(\tau)$ is usually referred to as the optimal force bias~\cite{Motta2018Sep} which provides a shift to the underlying normal distribution.
The optimal force bias is computed as
\begin{align}\nonumber
\bar{x}_{\gamma, i}(\tau) &= - \sqrt{\Delta \tau} \frac{\langle \Psi_{{T}} | \hat{v}_{\gamma}' |\psi_{i}{(\tau)} \rangle}{\langle \Psi_{{T}} |\psi_{i}{(\tau)} \rangle} \\
&=
-\sqrt{\Delta\tau} \sum_{pq}L_{pq}^\gamma G_{qp,i}(\tau),
\label{force_bias}
\end{align}
where
the one-body Green's function $G_{qp,i}(\tau)$ is given by
\begin{equation}
G_{qp,i}(\tau) = 
\frac{\langle \Psi_{{T}} | a_p^\dagger a_q |\psi_{i}{(\tau)} \rangle}{\langle \Psi_{{T}} |\psi_{i}{(\tau)} \rangle}.
\label{eq:1gf}
\end{equation}
Beyond~\cref{import_sampling}, ph-AFQMC imposes a constraint to ensure the positivity of the weights throughout the 
imaginary time propagation.
Such a constraint is achieved by
a modified importance function, $I_{\text{ph}}$,
\begin{align}\nonumber
I_{\text{ph}} ( \textbf{x}_i(\tau), \bar{\textbf{x}}_i(\tau), |\psi_{i}{(\tau)}\rangle ) 
&= 
||I ( \textbf{x}_i(\tau), \bar{\textbf{x}}_i(\tau), |\psi_{i}{(\tau)}\rangle )||\\
& \times \text{max}\left(0, \cos \theta_i (\tau) \right),
\label{mod_import_func}
\end{align}
where the phase $\theta_i(\tau)$ is given in \cref{eq:phase}.
This modified importance function is used in ph-AFQMC to update the weights.

\subsection{Numerical investigation of size-consistency}\label{appsubsec:size}

\begin{figure}[!ht]
    \centering
    \scalebox{0.45}{\includegraphics{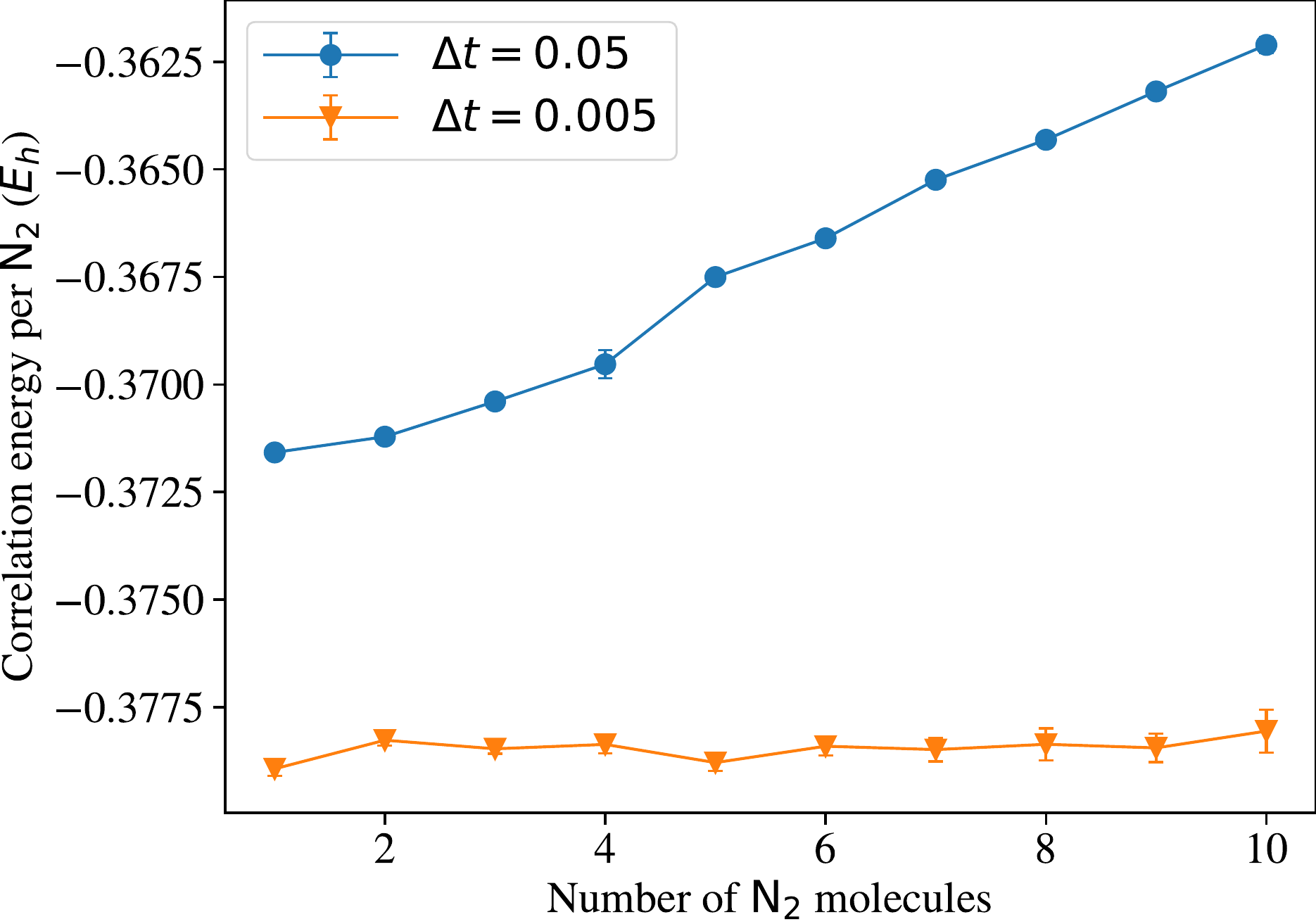}}
    \caption{Correlation energy per \ce{N2} molecule as a function of the number of \ce{N2} molecules that are infinitely far apart from each other for $\Delta t = 0.05$ a.u. (round, blue) and $\Delta t = 0.005$ a.u. (triangle, orange).}
    \label{fig:size}
\end{figure}

We verify the size-inconsistency error of ph-AFQMC with a finite time step via numerical results on \ce{N2} at a bond distance of 1.3 \AA~ and with an RHF trial wave function in the cc-pVDZ basis.\cite{Dunning1989Jan}
We add \ce{N2} molecules at infinite separation. One should expect the correlation energy per molecule to be constant if the method was size-consistent.
We used 640 walkers and checked that this result is not due to the population control bias. 
As shown in~\cref{fig:size}, for a relatively large time step ($\Delta t = 0.05$ a.u.), the correlation energy per molecule decreases as we add more \ce{N2} molecules to the system.
Going from one \ce{N2} molecule to 10 \ce{N2} molecules, we observe the change in the correlation energy per molecule by 10 m$E_h$ for $\Delta = 0.05$ a.u.
This confirms that ph-AFQMC is in general not size-consistent as illustrated by~\cref{eq:cosadd}.
We do not see such a large error in the case of a smaller time step ($\Delta t = 0.005$ a.u.). 
For larger systems, we anticipate that $\Delta t = 0.005$ a.u. would not be small enough to completely eliminate this error, and thus
one must be cautious with respect to time step and the possibility of size-inconsistency.
We also emphasize that when quantifying the size-inconsistency error in this numerical experiment, the typical time step error caused by the Trotter error in \cref{eq:Bx}
is the same for each fragment.
Therefore, our result is not due to the time step error during the Trotterization, and is solely caused by the lack of size-consistency in ph-AFQMC for finite $\Delta t$.

\subsection{Additional computational details concerning benchmark data}\label{appsub:compdetails}
We used 640 walkers in all systems considered here and the corresponding population control bias was found to be negligible with the pair-branch algorithm.~\cite{wagner_qwalk}
A time step of 0.005 au was used and the time step error (for both Trotterization and size-consistency) was found to be also negligible at the energy scale we focus on in this work.
The frozen core approximation was used throughout unless specified otherwise.

For the W4-11 set,~\cite{KARTON2011165} we used the aug-cc-pV5Z basis set~\cite{Dunning1989Jan} to converge the Hartree-Fock (HF) energies to the basis set limit and
used aug-cc-pVTZ~\cite{Dunning1989Jan} and aug-cc-pVQZ~\cite{Dunning1989Jan} to extrapolate the correlation energies to the basis set limit following Helgaker's two-point extrapolation.~\cite{Helgaker1997Jun}
We also use Karton and Martin's inner-shell correlation energy contribution to correct for the missing core correlation energy due to the frozen core approximation.~\cite{KARTON2011165}
Furthermore, to estimate the residual basis set error due to the differences in our scheme and Karton and Martin's scheme, we also compared our CCSD(T) relative energies
and their reported CCSD(T) relative energies. The root-mean-square-deviation of the difference was 0.66 kcal/mol which is small enough for the purpose of this paper.
We note that Karton and Martin used the composite W4 scheme to estimate reference energies which involved up to CCSDTQ5 or CCSDTQ6.
For the A24 set,~\cite{rezac2013describing} we used aug-cc-pVTZ\cite{Dunning1989Jan} with counterpoise corrections, although the reference data (i.e., CCSD with triples and perturbative quadruples) were obtained in the basis set limit.
For \ce{H4} and \ce{N2} dissociation data, we used cc-pVQZ and cc-pVTZ, respectively,  bases\cite{Dunning1989Jan} without the frozen core approximation.

Molecular integrals for ph-AFQMC were generated by PySCF,~\cite{Sun2020Jul}
all deterministic single-reference quantum chemistry calculations were performed with Q-Chem,~\cite{Epifanovsky2021Aug}
and
AFQMC calculations were performed with
the ipie~\cite{ipie,Malone2022Sep} and QMCPACK.~\cite{Kent2020May}
Cholesky factorization and integral transformations necessary for AFQMC calculations were performed by a script in ipie~\cite{ipie,Malone2022Sep} with a Cholesky threshold of $10^{-5}$.
Semistochastic heat-bath configuration interaction (SHCI) calculations were performed with Dice.~\cite{dice}
Multi-reference dynamic correlation methods other than driven similarity renormalization group (DSRG) methods were run on Orca.~\cite{Neese2018Jan}
DSRG methods were run through Psi4 with the flow parameter $\sigma = 0.5$ a.u..~\cite{Smith2020May}
DSRG-MRPT2 and DSRG-MRPT3 energies were obtained via the partial relaxation scheme.~\cite{Li2017Mar}
MR-LDSRG(2) energies were obtained by relaxing the reference state twice.~\cite{Li2016Apr}

We used SHCI to generate near-exact benchmark energies for bond dissociation examples.
SHCI energies are converged more accurately than 0.1 m$E_h$, except for \ce{N2} in the cc-pVTZ basis.
For \ce{N2} in the cc-pVTZ basis, we estimate the remaining bias in our SHCI energies to be less than 1 kcal/mol.
The second-order perturbation theory (PT2) contribution was found to be at most 10 m$E_h$ for this system and we did not perform any extrapolation to reduce the potential bias further. 
The SHCI energies used in this work are within 1 m$E_h$ of 
the SHCI energies from our recent investigation of \ce{N2} in the cc-pVTZ basis at a limited set of bond distances~\cite{Huggins2022Mar} where the PT2 contribution in SHCI was converged to better than 2 m$E_h$.
We believe that the residual bias in SHCI energies is small enough that our conclusions are not affected by this bias.
$\omega$B97M-V results are taken from ref. \citenum{Mardirossian2017}

All relevant raw energies are available in the Zenodo repository.\cite{zenodo}

\subsection{Potential energy curves of \ce{H4} and \ce{N2} in STO-3G}\label{subsec:pessto3g}
\begin{figure}[!h]
    \centering
    \scalebox{0.38}{\includegraphics{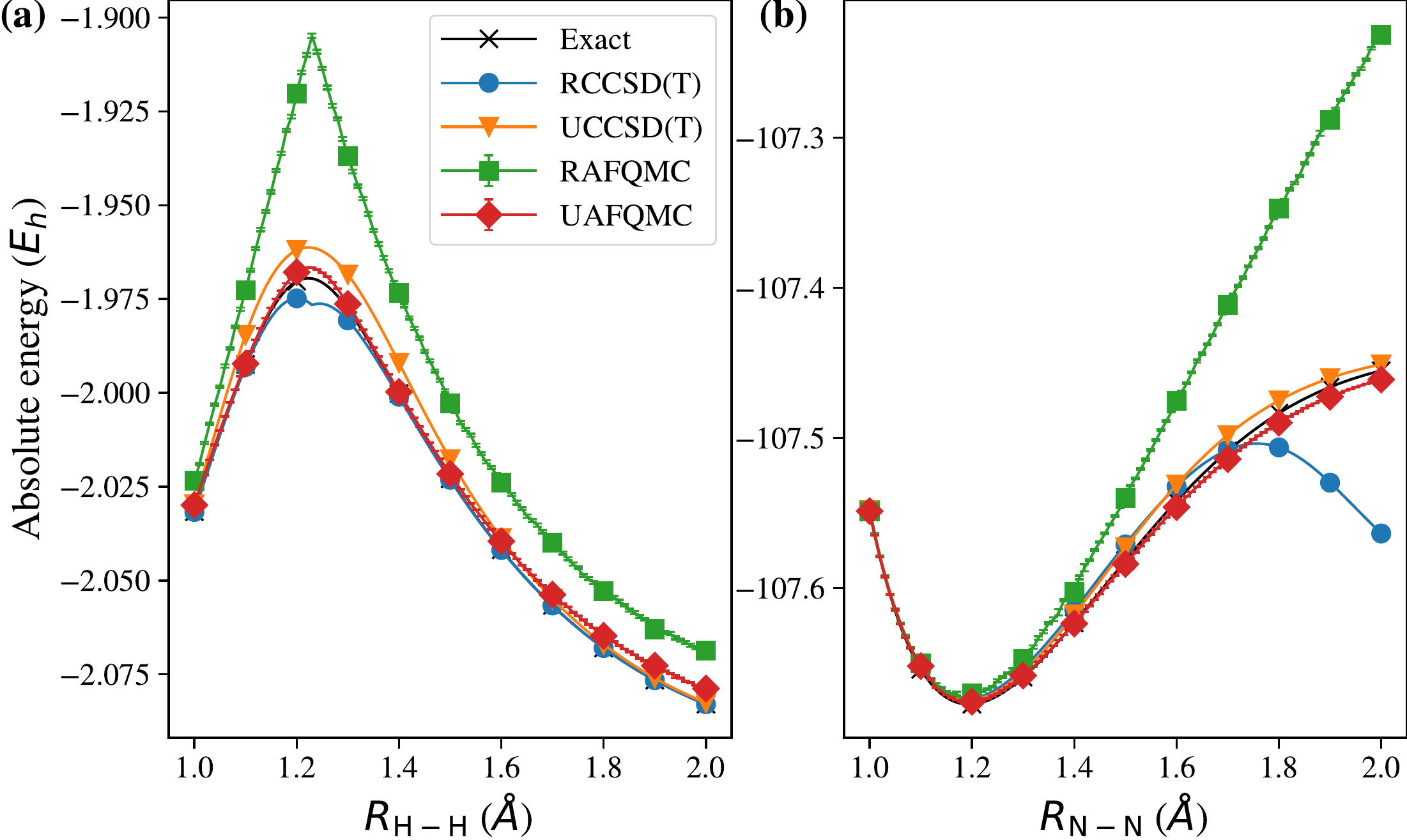}}
    \caption{The potential energy curve of (a) \ce{H4} and (b) \ce{N2} as a function of the bond distance using an STO-3G basis set.
        Note that there are more data points than markers shown.}
    \label{fig:pesabs}
\end{figure}

\begin{figure}[!h]
    \centering
    \scalebox{0.38}{\includegraphics{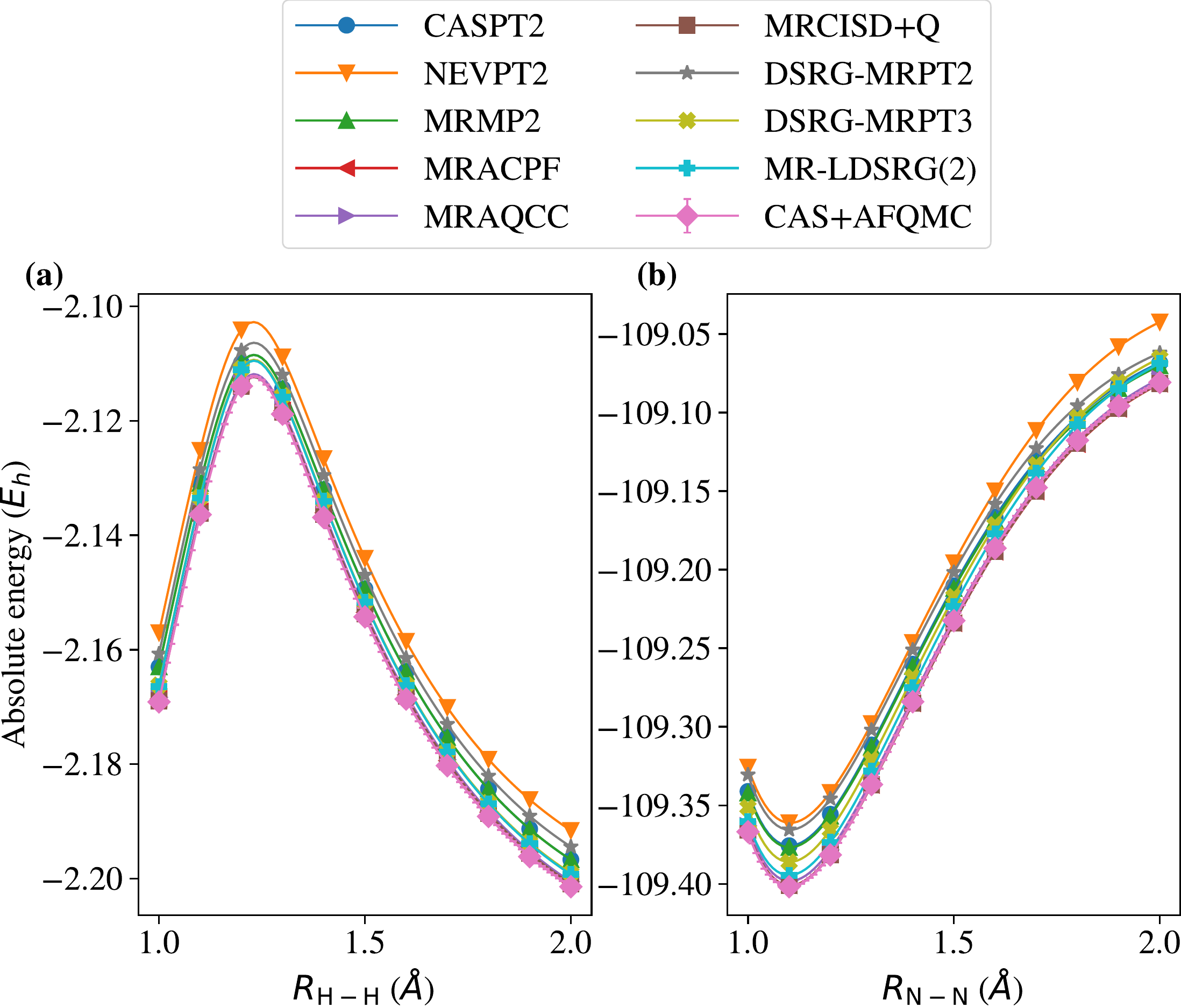}}
    \caption{The potential energy curve of (a) \ce{H4} in cc-pVQZ and (b) \ce{N2} in cc-pVTZ as a function of the bond distance using multi-reference methods..
        Note that there are more data points than markers shown.}
    \label{fig:pesabsmr}
\end{figure}

We present the potential energy curve of \ce{H4} and \ce{N2} using an STO-3G basis set with single-reference methods in \cref{fig:pesabs} and multi-reference methods in \cref{fig:pesabsmr}.
This is to give the reader a visual sense for the potential energy curves for each of these problems.

\bibliography{refs}

\end{document}